\documentclass[10pt]{bmc_article}    

\usepackage{cite} 
\usepackage{url}  
\usepackage{ifthen}  
\usepackage{multicol}   
\usepackage[utf8]{inputenc} 
\usepackage{pgf} 
\usepackage{amsfonts}

\def\includegraphics{}

\newboolean{publ}

\newenvironment{bmcformat}{\begin{raggedright}\baselineskip20pt\sloppy\setboolean{publ}{false}}{\end{raggedright}\baselineskip20pt\sloppy}

\def \hc{{\sc HMM-Converter}}

\begin{document}
\begin{bmcformat}


\title{Efficient algorithms for training the parameters of hidden Markov models using stochastic expectation maximization (EM)~training and 
Viterbi training}
\author{
        Tin Y~Lam\,$^{1}$\email{Tin Y.~Lam - natural@cs.ubc.ca} \and
        Irmtraud M~Meyer\correspondingauthor,$^{1}$\email{Irmtraud M.~Meyer - irmtraud.meyer@cantab.net}}

\address{\iid(1) Centre for High-Throughput Biology, Department of
        Computer Science and Department of Medical Genetics, 2366 Main Mall,
        University of British Columbia, Vancouver V6T 1Z4, Canada}

\maketitle

\begin{abstract}
\paragraph{Background:}

Hidden Markov models are widely employed by numerous bioinformatics programs
used today. Applications range widely from comparative gene prediction to
time-series analyses of micro-array data. The parameters of the underlying
models need to be adjusted for specific data sets, for example the genome of a
particular species, in order to maximize the prediction
accuracy. Computationally efficient algorithms for parameter training are thus
key to maximizing the usability of a wide range of bioinformatics applications.

\paragraph{Results:}
We introduce two computationally efficient training algorithms, one for
Viterbi training and one for stochastic expectation maximization (EM)
training, which render the memory requirements independent of the sequence
length. Unlike the existing algorithms for Viterbi and stochastic EM~training
which require a two-step procedure, our two new algorithms require only one
step and scan the input sequence in only one direction. We also implement
these two new algorithms and the already published linear-memory algorithm for
EM~training into the hidden Markov model compiler \hc\ and examine their
respective practical merits for three small example models.


\paragraph{Conclusions:}
Bioinformatics applications employing hidden Markov models can use the two
algorithms in order to make Viterbi training and stochastic EM~training more
computationally efficient. Using these algorithms, parameter training can thus
be attempted for more complex models and longer training sequences. The two
new algorithms have the added advantage of being easier to implement than the
corresponding default algorithms for Viterbi training and stochastic
EM~training.

\end{abstract}

\ifthenelse{\boolean{publ}}{\begin{multicols}{2}}{}

\section*{Background}


Hidden Markov models (HMMs) and their variants are widely used for analyzing
biological sequence data. Bioinformatics applications range from methods for
comparative gene prediction~(e.g.\ \cite{Meyer2004, Stanke2006b}) to methods
for modeling promoter grammars~(e.g.\ \cite{Won2008}), identifying protein
domains~(e.g.\ \cite{Finn2008}), predicting protein interfaces~(e.g.\
\cite{Nguyen2007}), the topology of transmembrane proteins~(e.g.\
\cite{Krogh2001}) and residue-residue contacts in protein structures~(e.g.\
\cite{Bjorkholm2009}), querying pathways in protein interaction
networks~(e.g.\ \cite{Qian2008}), predicting the occupancy of transcription
factors~(e.g.\ \cite{Drawid2009}) as well as inference models for genome-wide
association studies~(e.g.\ \cite{Hosking2008}) and disease association tests
for inferring ancestral haplotypes~(e.g.\ \cite{Su2008}).


Most of these bioinformatics applications have been set up for a specific type
of analysis and a specific biological data set, at least initially. The states
of the underlying HMM and the implemented prediction algorithms determine
which type of data analysis can be performed, whereas the parameter values of
the HMM are chosen for a particular data set in order to optimize the
corresponding prediction accuracy. If we want to apply the same method to a
new data set, e.g.\ predict genes in a different genome, we need to adjust the
parameter values in order to make sure the performance accuracy is optimal.


Manually adjusting the parameters of an HMM in order to get a high prediction
accuracy can be a very time consuming task which is also not guaranteed to
improve the performance accuracy. A variety of training algorithms have
therefore been devised in order to address this challenge. These training
algorithms require as input and starting point a so-called \emph{training set}
of (typically partly annotated) data. Starting with a set of (typically
user-chosen) initial parameter values, the training algorithm employs an
iterative procedure which subsequently derives new, more refined parameter
values. The iterations are stopped when a termination criterion is met, e.g.\
when a maximum number of iterations have been completed or when the change of
the log-likelihood from one iteration to the next become sufficiently
small. The model with the final set of parameters is then used to test if the
performance accuracy has been improved. This is typically done by analyzing a
\emph{test set} of annotated data which has no overlap with the training set
by comparing the predicted to the known annotation.


Of the training algorithms used in bioinformatics applications, the Viterbi
training algorithm~\cite{Juang1990, Durbin1998} is probably the most commonly
used, see e.g.\ \cite{Besemer2001, Lunter2007, Ter-Hovhannisyan2008}.  This is
due to the fact that it is easy to implement if the Viterbi
algorithm~\cite{viterbi1967} is used for generating predictions. In each
iteration of Viterbi training, a new set of parameter values $\phi$ is derived
from the counts of emissions and transitions in the Viterbi paths $\Pi^*$ for
the set of training sequences $\mathcal{X}$. Because the new parameters are
completely determined by the Viterbi paths, Viterbi training converges as soon
as the Viterbi paths no longer change or, alternatively, if a fixed number of
iterations have been completed. Viterbi training finds at best a local optimum
of the likelihood $P(\mathcal{X}, \Pi^*|\phi)$, i.e.\ it derives parameter
values $\phi$ that maximize the contribution from the set of Viterbi paths
$\Pi^*$ to the likelihood. There already exist a number of algorithms that can
make Viterbi decoding computationally more efficient.  Keibler \emph{et
  al.}~\cite{Keibler2007} introduce two heuristic algorithms for Viterbi
decoding which they implement into the gene-prediction program {\sc
  Twinscan/N-SCAN}, called ``Treeterbi'' and ``Parallel Treeterbi'', which
have the same worst case asymptotic memory and time requirements as the
standard Viterbi algorithm, but which in practice work in a significantly more
memory efficient way. Sramek \emph{et al.}~\cite{Sramek2007} present a new
algorithm, called ``on-line Viterbi algorithm'' which renders Viterbi decoding
more memory efficient without significantly increasing the time
requirement. The most recent contribution is from Lifshits \emph{et
  al.}~\cite{Lifshits2009} who propose more efficient algorithms for Viterbi
decoding and Viterbi training. These new algorithms exploit repetitions in the
input sequences (in five different ways) in order to accelerate the default
algorithm.

Another well-known training algorithm for HMMs is Baum-Welch
training~\cite{baum1972} which is an expectation maximization (EM)
algorithm~\cite{dlr1977}. In each iteration, a new set of parameter values is
derived from the estimated number of counts of emissions and transitions by
considering \emph{all} possible state paths (rather than only a single Viterbi
path) for every training sequence. The iterations are typically stopped after
a fixed number of iterations or as soon as the change in the log-likelihood is
sufficiently small.  For Baum-Welch training, the likelihood
$P(\mathcal{X}|\phi)$~\cite{Durbin1998} can be shown to converge (under some
conditions) to a stationary point which is either a local optimum or a saddle
point. Baum-Welch training using the traditional combination of forward and
backward algorithm~\cite{Durbin1998} is, for example, implemented into the
prokaryotic gene prediction method {\sc EasyGene}~\cite{Larsen2003} and the
HMM-compiler {\sc HMMoC}~\cite{Lunter2007}. As for Viterbi training, the
outcome of Baum-Welch training may strongly depend on the chosen set of
initial parameter values. As Jensen~\cite{Jensen2009} and Khreich \emph{et
  al.}~\cite{Khreich2010} describe, computationally more efficient algorithms
for Baum-Welch training which render the memory requirement independent of the
sequence length have been proposed, first in the communication field by
\cite{Elliott1995, Sivaprakasam1995, Turin1998} and later, independently, in
bioinformatics by Mikl\'os and Meyer~\cite{Miklos2005}, see also
\cite{Churbanov2008}. The advantage of this linear-memory memory algorithm is
that it is comparatively easy to implement as it requires only a one- rather
than a two-step procedure and as it scans the sequence in a uni- rather than
bi-directional way. This algorithm was employed by Hobolth and
Jensen~\cite{Hobolth2005} for comparative gene prediction and has also been
implemented, albeit in a modified version, by Churbanov and
Winters-Hilt~\cite{Churbanov2008} who also compare it to other implementations
of Viterbi and Baum-Welch training including checkpointing implementations.

Stochastic expectation maximization (EM)~training or Monte Carlo EM training
~\cite{Bishop2006} is another iterative procedure for training the parameters
of HMMs. Instead of considering only a \emph{single} Viterbi~state path for a
given training sequence as in Viterbi training or \emph{all} state paths as in
Baum-Welch training, stochastic EM~training considers a fixed-number of $K$
state paths $\Pi^s$ which are sampled from the posterior distribution
$P(\Pi|X)$ for every training sequence $X$ in every iteration. Sampled state
paths have already been used in several bioinformatics applications for
sequence decoding, see e.g.\ \cite{Cawley2003, Stanke2006b} where sampled
state paths are used in the context of gene prediction to detect alternative
splice variants.

All three above training algorithms, i.e.\ Viterbi training, Baum-Welch
training and stochastic EM~training, can be combined with the traditional
check-pointing algorithm~\cite{Grice1997, Tarnas1998, Wheeler2000} in order to
trade time for memory requirements.


We here introduce two new algorithms that make Viterbi training and stochastic
EM~training computationally more efficient. Both algorithms have the
significant advantage of rendering the memory requirement independent of the
sequence length for HMMs while keeping the time requirement the same (for
Viterbi training) or modifying it by a factor of $M K /(M +K)$, i.e.\
decreasing it when only one state path $K = 1$ is sampled for a model of $M$
states (for stochastic EM~training). Both algorithms are inspired by the
linear-memory algorithm for Baum-Welch training which requires only a
uni-directional rather than bi-directional movement along the input sequence
and which has the added advantage of being considerably easier to
implement. We present a detailed description of the two new algorithms for
Viterbi training and stochastic EM~training. In addition, we implement all
three algorithms, i.e.\ the new algorithms for Viterbi training and stochastic
EM~training and the previously published linear-memory algorithm for
Baum-Welch training, into our HMM-compiler \hc~\cite{Lam2009} and examine the
practical features of these these three algorithms for three small example
HMMs.

\section*{Methods and Results}

\subsection*{Definitions and notation}


In order to simplify the notation in the following, we will assume without
loss of generality that we are dealing with a 1st-order HMM where the
$\textsl{Start}$ state and the $\textsl{End}$ state are the only silent
states. Our description of the existing and the new algorithms easily
generalize to higher-order HMMs, HMMs with more silent states (provided there
exists no circular path in the HMM involving only silent states) and $n$-HMMs,
i.e.\ HMMs which read $n$ un-aligned input sequences rather than a single
input sequence at a time.

An HMM is defined by
\begin{itemize}
\item a set of states $\mathcal{S} = \{0, 1, \ldots, M\}$, where state $0$ denotes the $\textsl{Start}$ 
      and state $M$ denotes the $\textsl{End}$ state and where all other states are non-silent, 
\item a set of transition probabilities $\mathcal{T} = \{t_{i,j} | i,j \in \mathcal{S} \}$,
      where $t_{i,j}$ denotes the transition probability to go from state $i$ to state $j$ and
      $\sum_{j \in \mathcal{S}} t_{i,j} = 1$ for every state $i \in \mathcal{S}$ and 
\item a set of emission probabilities $\mathcal{E} = \{e_i(y) | i \in \mathcal{S}, 
      y \in \mathcal{A} \}$, where $e_i(y)$ denotes the emission probability of state $i$ for
      symbol $y$ and $\sum_{y \in \mathcal{A}}e_i(y) = 1$ for every non-silent state $i \in \mathcal{S}$
      and $\mathcal{A}$ denotes the alphabet from which the symbols in the input sequences are derived, 
      e.g.\ $\mathcal{A} = \{ {\tt A}, {\tt C}, {\tt G}, {\tt T}\}$ when dealing with DNA~sequences.
\end{itemize}

We also define:
\begin{itemize}
\item $T_{max}$ is the maximum number of states that any state in the model is connected to, also called the model's connectivity.
\item $\mathcal{X} = \{X^1, X^2, \ldots, X^N\}$ denotes the training set of $N$ sequences, where each particular 
      training sequence $X^i$ of length $L^i$ is denoted $X^i = (x^i_1, x^i_2, \ldots, x^i_{L^i})$. 
      In the following and to simplify the notation, we pick one particular training
      sequence $X \in \mathcal{X}$ of length $L$ as representative which we denote $X = (x_1, x_2, \ldots, x_{L})$. 
      We write $X_n = (x_1, x_2, \ldots, x_n)$, $n \in \{1, \ldots, L\}$, to denote the sub-sequence of $X$ which finishes at 
      sequence position $n$.
\item $\Pi = (\pi_0, \pi_1, \ldots, \pi_{L+1})$ denotes a state path in the HMM for an input sequence $X$ of length 
      $L$, i.e.\ state $\pi_i$ is assigned to sequence position $x_i$. $\Pi^*$ denotes a Viterbi path and $\Pi^s$ a state path that has
      been sampled from the posterior distribution $P(\Pi|X)$ of the corresponding sequence $X$.
\end{itemize}

\subsection*{A linear-memory algorithm for Viterbi training}

Of the HMM-based methods that provide automatic algorithms for parameter
training, Viterbi training~\cite{Durbin1998} is the most popular. This is
primarily due to the fact that Viterbi training is readily implemented if the
Viterbi algorithm is used to generate predictions.  Similar to Baum-Welch
training~\cite{baum1972, dlr1977}, Viterbi training is an iterative training
procedure. Unlike Baum-Welch training, however, which considers \emph{all}
state paths for a given training sequence in each iteration, Viterbi training
only considers a \emph{single} state path, namely a Viterbi path, when
deriving new sets of parameters. In each iteration, a new set of parameter
values is derived from the counts of emissions and transitions in the Viterbi
paths~\cite{viterbi1967} of the training sequences. The iterations are
terminated as soon as the Viterbi paths of the training sequences no longer
change.

In the following, 

\begin{itemize}
\item let $E^q_i(y, X, \Pi^*(X))$ denote the number of times that state $i$ reads symbol $y$ from 
      input sequence $X$ in Viterbi path $\Pi^*(X)$ given the HMM with parameters from the $q$-th iteration,
\item in particular let $E^q_i(y, X_k, \Pi^*(X_k, \pi^*_k=m ))$ denote the number of times that state $i$ reads symbol $y$ from 
      input sequence $X$ in the partial Viterbi path $\Pi^*(X_k, \pi^*_k=m) = (\pi^*_0, \ldots, \pi^*_{k-1}, \pi^*_k = m)$ 
      which finishes at sequence position $k$ in state $m$, and
\item let $T^q_{i,j}(X, \Pi^*(X))$ denote the number of times that a transition from state $i$ to state $j$ is
      used in Viterbi path $\Pi^*(X)$ for sequence $X$ given the HMM with parameters from the $q$-th iteration,
\item in particular let $T^q_{i,j}(X_k, \Pi^*(X_k, \pi^*_k = m))$ denote the number of times that a transition from state $i$ to state $j$ is
      used in the partial Viterbi path $\Pi^*(X_k, \pi^*_k=m) = (\pi^*_0, \ldots, \pi^*_{k-1}, \pi^*_k = m)$ 
      which finishes at sequence position $k$ in state $m$.
\end{itemize}

In the following, the superscript $q$ will indicate from which iteration the
underlying parameters derive. If we consider all $N$ sequences of a training
set $\mathcal{X} = \{X^1, \ldots X^N\}$ and a Viterbi path $\Pi^*(X^n)$ for each
sequence $X^n$ in the training set, the recursion which updates the values of
the transition and emission probabilities reads:

\begin{eqnarray}
  t^{q+1}_{i,j} & = & \frac{\sum_{n=1}^N  T^q_{i,j}(X^n, \Pi^*(X^n))}
                         {\sum_{j=1}^M \sum_{n=1}^N T^q_{i,j}(X^n, \Pi^*(X^n))} \label{eq:T_update} \\ 
  e^{q+1}_i(y) & = & \frac{\sum_{n=1}^N E^q_i(y, X^n, \Pi^*(X^n))}
                         {\sum_{y' \in \mathcal{A}} \sum_{n=1}^N E^q_i(y', X^n, \Pi^*(X^n))} \label{eq:E_update} 
\end{eqnarray}

These equations assume that we know the values of $T^q_{i,j}(X^n, \Pi^*(X^n))$
and $E^q_i(y, X^n, \Pi^*(X^n))$, i.e.\ how often each transition and emission
is used in the Viterbi path $\Pi^*(X^n)$ for training sequence $X^n$.

One straightforward way to determine $T^q_{i,j}(X^n, \Pi^*(X^n))$ and $E^q_i(y,
X^n, \Pi^*(X^n))$ is to first calculate the two-dimensional Viterbi matrix for
every training sequence $X^n$, to then derive a Viterbi state path $\Pi^*(X^n)$
from each Viterbi matrix using the well-known traceback
procedure~\cite{viterbi1967} and to then simply count how often each
transition and each emission was used. Using this strategy, every iteration in
the Viterbi training algorithm would require $\mathcal{O}(M \max_i\{L_i\} +
\max_i\{L_i\})$ memory and $\mathcal{O}(M T_{max} \sum_{i=1}^N L_i +
\sum_{i=1}^N L_i)$ time, where $\sum_{i=1}^N L_i$ is the sum of the $N$
sequence lengths in the training set $\mathcal{X}$ and $\max_i\{L_i\}$ the
length of the longest sequence in training set $\mathcal{X}$.  However, for
many bioinformatics applications where the number of states in the model $M$
is large, the connectivity $T_{max}$ of the model high or the training
sequences are long, these memory and time requirements are too large to allow
automatic parameter training using this algorithm.

A linear-memory version of the Viterbi algorithm, called the Hirschberg
algorithm~\cite{Hirschberg1975}, has been known since 1975. It can be used to
derive Viterbi paths in memory that is linearized with respect to the length
of one of the input sequences while increasing the time requirement by at most
a factor of two. The Hirschberg algorithm, however, only applies to $n$-HMMs
with $n \geq 2$, i.e.\ HMMs which read two or more un-aligned input sequences
at a time. One significant disadvantage of the Hirschberg algorithm is that it
is considerably more difficult to implement than the Viterbi algorithm. Only
few HMM-based applications in bioinformatics actually employ it, see e.g.\
\cite{Meyer02, Meyer2004, Lam2009}. We will see in the following how we can
devise a linear-memory algorithm for Viterbi training that does not involve
the Hirschberg algorithm and that can be applied to all $n$-HMMs including
$n=1$.


We now introduce a linear-memory algorithm for Viterbi training.  The idea for
this algorithm stems from the following observations:

\begin{itemize}
\item[(V1)] If we consider the description of the Viterbi
  algorithm~\cite{viterbi1967}, in particular the recursion, we realize that
  the calculation of the Viterbi values can be continued by retaining only the
  values for the previous sequence position.

\item [(V2)] If we have a close look at the description of the traceback
  procedure~\cite{viterbi1967}, we realize that we only have to remember the
  Viterbi matrix elements at the \emph{previous} sequence position in order to
  deduce the state from which the Viterbi matrix element at the \emph{current}
  sequence position and state was derived.

\item[(V3)] If we want to derive the Viterbi path $\Pi$ from the Viterbi matrix,
  we have to start at the end of the sequence in the $\textsl{End}$ state $M$. 
\end{itemize}

Observations~(V1) and (V2) imply that local information suffices to continue
the calculation of the Viterbi matrix elements (V1) and to derive a previous
state (V2) if we already are in a particular state and sequence position,
whereas observation~(V3) reminds us that in order to derive the Viterbi path,
we have to start at the \emph{end} of the training sequence. Given these three
observations, it is not obvious how we can come up with a computationally more
efficient algorithm for training with Viterbi paths. In order to realize that
a more efficient algorithm exists, one also has to also note that:

\begin{itemize}
\item[(V4)] While calculating the Viterbi matrix elements in the
  memory-efficient way outlined in (V1), we can \emph{simultaneously} keep
  track of the previous state from which the Viterbi matrix element at every
  current state and sequence position was derived. This is possible because of
  observation (V2) above.

\item[(V5)] In every iteration $q$ of the training procedure, we only need to
  know the values of $T^q_{i,j}(X, \Pi^*(X))$ and $E^q_i(y, X, \Pi^*(X))$, i.e.\ how
  \emph{often} each transition and emission was used in each Viterbi state
  path $\Pi^*(X)$ for every training sequence $X$, but not
  \emph{where} in the Viterbi matrix each transition and emission was
  used.
\end{itemize}

Given all observations~(V1) to (V5), we can now formally write down an
algorithm which calculates $T^q_{i,j}(X, \Pi^*(X))$ and $E^q_i(y, X,
\Pi^*(X))$ in a computationally efficient way which linearizes the memory
requirement with respect to the sequence length and which is also easy to
implement. In order to simplify the notation, we describe the following
algorithm for one particular training sequence $X$ and omit the superscript for
the iteration $q$, as both remain the same throughout the algorithm. In the
following,

\begin{itemize}
\item $T_{i,j}(k, m)$ denotes the number of times the transition from state
$i$ to state $j$ is used in a Viterbi state path that finishes at sequence
position $k$ in state $m$,
\item $E_i(y, k, m)$ denotes the number of times that state $i$ reads symbol
  $y$ in a Viterbi state path that finishes at sequence position $k$ in state
  $m$,
\item $v_i(k)$ denotes the Viterbi matrix element for state $i$ and sequence
  position $k$, i.e.\ $v_i(k)$ is the probability of the Viterbi state path,
  i.e.\ the state path with the highest overall probability, that starts at
  the beginning of the sequence in the $\textsl{Start}$ state and finishes in
  state $i$ as sequence position $k$, 
\item $i,j,n \in \mathcal{S}$, $y \in \mathcal{A}$ and $l\in \mathcal{S}$
denotes the previous state from which the current Viterbi matrix element
$v_m(k)$ was derived, and 
\item $\delta_{i,j}$ is the delta-function with $\delta_{i,j} = 1$ for $i=j$
  and $\delta_{i,j} = 0$ else.
\end{itemize}

{\bf Initialization}: at the start of training sequence $X = (x_1, \ldots,
x_L) $ and for all $m \in \mathcal{S}$, set
\begin{eqnarray}
v_m(0) & = & 
\left\{
\begin{array}{rcl}
1 & m = 0  \nonumber\\
0 & m \neq 0
\end{array}
\right. \nonumber \\
T_{i,j} (0, m) & = & 0 \nonumber \\
E_i(y, 0, m) &=& 0 \nonumber
\end{eqnarray}

{\bf Recursion}: loop over all positions $k$ from 1 to $L$ in the training
sequence $X$ and loop, for each such sequence position $k$, 
over all states $m \in \mathcal{S} \backslash \{0\} = \{1, \ldots, M\}$ and set
\begin{eqnarray}
v_m(k)         & = & e_m(x_{k}) \cdot \max_{n \in \mathcal{S}} \left\{ v_n(k-1)\cdot t_{n,m} \right\}        \nonumber \\
T_{i,j}(k, m)   & = & T_{i,j}(k-1, l) + \delta_{l,i} \cdot \delta_{m,j}          \nonumber \\
E_i(y, k, m)   & = & E_i(y, k-1, l) + \delta_{m,i} \cdot \delta_{y, x_k}          \nonumber
\end{eqnarray}

where $l$ denotes the state at the previous sequence position $k-1$ from
which the Viterbi matrix element $v_m(k)$ for state $m$ and sequence position
$k$ derives, i.e.\ $l = \textrm{argmax}_{n \in \mathcal{S}} \left\{ v_n(k-1)\cdot t_{n,m} \right\}$.

{\bf Termination}: at the end of the input sequence, i.e.\ for $k = L$ and for $m = M$
the silent \emph{End} state, set
\begin{eqnarray}
v_{M}(L)      & = & \max_{n \in \mathcal{S}} \left\{ v_n(L)\cdot t_{n,M} \right\}        \nonumber \\
T_{i,j}(L, M) & = & T_{i,j}(L, l) + \delta_{l,i} \cdot \delta_{M,j}             \nonumber \\
E_i(y, L, M)  & = & E_i(y, L, l)                                            \nonumber
\end{eqnarray}

where $l$ denotes the state at the sequence position $L$ from which the
Viterbi matrix element $v_M(L)$ for the \emph{End} state $M$ and sequence
position $L$ derives, i.e.\ $l = \textrm{argmax}_{n \in \mathcal{S}} \left\{ v_n(L)\cdot t_{n,M}
\right\}$.

The above algorithm yields $T_{i,j}(L,M) = T_{i,j}^q(X, \Pi^*(X))$ and $E_i(y,
L, M) = E_i^q(y, X, \Pi^*(X))$ (and $v_M(L) = P^q(X, \Pi^*(X))$), i.e.\ we
know how often a transition from state $i$ to state $j$ was used and how often
symbol $y$ was read by state $i$ in Viterbi state path $\Pi^*(X)$ in iteration
$q$.

{\bf Theorem 1:} The above algorithm yields $T_{i,j}(L,M) = T_{i,j}^q(X,
\Pi^*(X))$ and $E_i(y, L, M) = E_i^q(y, X, \Pi^*(X))$.

{\bf Proof:} We will prove these statements via induction with respect to the
sequence position $k$.

{\bf (1) Induction start at $k=0$:} This corresponds to the initialization
step in the algorithm. $T_{i,j}(0,m)=0$ and $E_i(y,0,m)=0$ for all $m \in
{\mathcal S}$ as any zero-length Viterbi path finishing in state $m$ at
sequence position $0$ has zero transitions from state $i$ to $j$ and has not
read any sequence symbol.

{\bf (2) Induction step $k-1 \rightarrow k$ for $k \in \{1, \ldots L-1\}$ if
  the state at sequence position $k=L$ is not the \emph{End} state $M$:} This
case corresponds to the recursion in the algorithm. We assume that
$T_{i,j}(k-1, m) = T_{i,j}^q(X_{k-1}, \Pi^*(X_{k-1}, \pi^*_{k-1} = m))$ and
$E_i(y, k-1, m) = E_i^q(y, X_{k-1}, \Pi^*(X_{k-1}, \pi^*_{k-1} = m))$.  We
need to distinguish two cases (a) and (b). Let $l$ denote the state at
sequence position $k-1$ from which the Viterbi matrix element $v_m(k)$ for
state $m$ and sequence position $k$ derives, i.e.\ $l = \textrm{argmax}_{n \in
  \mathcal{S}} \left\{ v_n(k-1)\cdot t_{n,m} \right\}$.

\begin{itemize}
\item {\bf Case (a):} 

{\bf Emissions (i): $m=i$ and $y=x_k$:} In this case, $E_i(y,k,m) =
E_i(y,k-1,l) + 1$. As we know that $E_i(y,k-1,l)$ is the number of times that
state $i$ reads symbol $y$ in a Viterbi path ending in state $l$ at sequence
position $k-1$, we need to add $1$ count for reading symbol $y=x_k$ by state
$m=i$ at the next sequence position $k$ in order to obtain $E_i(y,k,m)$.

{\bf Transitions (ii): $l=i$ and $m=j$:} In this case, $T_{i,j}(k, m) =
T_{i,j}(k-1, l) + 1$. As we know that $T_{i,j}(k-1, l)$ is the number of times
that a transition from state $i$ to state $j$ is used in a Viterbi path ending
in state $l$ at sequence position $k-1$, we need to add $1$ count for the
transition from state $l=i$ to state $m=j$ which brings us from sequence
position $k-1$ to $k$ in order to get $T_{i,j}(k, m)$.

\item {\bf Case (b):} 

{\bf Emissions (i): $m \ne i$ or $y \ne x_k$:} In this case, $E_i(y,k,m) =
E_i(y,k-1,l)$. We know that $E_i(y,k-1,l)$ is the number of times that state
$i$ reads symbol $y$ in a Viterbi path ending in state $l$ at sequence
position $k-1$. If we go from state $l$ at position $k-1$ to state $m$ at
position $k$ and read symbol $x_k$ and if $m \ne i$ or $y \ne x_k$, we do not
need to modify the number of counts as we know that state $i$ at position $k$
does not read symbol $y$, i.e.\ $E_i(y,k,m) = E_i(y,k-1,l)$.

{\bf Transitions (ii): $l \ne i$ or $m \ne j$:} In this case, $T_{i,j}(k, m) =
T_{i,j}(k-1, l)$. We know that $T_{i,j}(k-1, l)$ is the number of times that a
transition from state $i$ to state $j$ is used in a Viterbi path ending in
state $l$ at sequence position $k-1$. If we make a transition from state $l$
at position $k-1$ to state $m$ at position $k$ and if $l \ne i$ or $m \ne j$,
we do not need to modify the number of counts as we know this is not a
transition from state $i$ to state $j$, i.e.\ $T_{i,j}(k, m) = T_{i,j}(k-1,
l)$.
\end{itemize}

{\bf (3) If the state at sequence position $k=L$ is the \emph{End} state $M$:}
This case corresponds to the termination step in the algorithm. As in (2), we
need to distinguish two cases (a) and (b), but now only for the transition
counts. Let $l$ denote the state at sequence position $L$ from which the
Viterbi matrix element $v_M(L)$ for the \emph{End} state $M$ and sequence
position $L$ derives, i.e.\ $l = \textrm{argmax}_{n \in \mathcal{S}} \left\{ v_n(L)\cdot t_{n,M}
\right\}$.

{\bf Emissions (i):} In this case, $E_i(y,L,M) = E_i(y,L,l)$. As we know that
$E_i(y,L,l)$ is the number of times that state $i$ reads symbol $y$ in a
Viterbi path ending in state $l$ at sequence position $L$, we do not need to
modify this number of counts when going to the silent \emph{End} state at the
same sequence position $L$ as silent states do not read any symbols from the
input sequence. As we are now at the end of the input sequence $X$ and the
Viterbi path $\Pi^*(X)$, we have $E_i(y, L, M) = E_i^q(y, X, \Pi^*(X))$.

\begin{itemize}
\item {\bf Case (a):} 

{\bf Transitions (i): $l=i$ and $M=j$:} In this case, $T_{i,j}(L, M) =
T_{i,j}(L, l) + 1$. As we know that $T_{i,j}(L, l)$ is the number of times
that a transition from state $i$ to state $j$ is used in a Viterbi path ending
in state $l$ at sequence position $L$, we need to add $1$ count for the
transition from state $l=i$ to the \emph{End} state $M=j$ at sequence position
$L$. Note that this transition of state does not incur a change of
sequence position as the \emph{End} state is a silent state.  As we are now at
the end of the input sequence $X$ and the Viterbi path $\Pi^*(X)$, we have
$T_{i,j}(L,M) = T_{i,j}^q(X, \Pi^*(X))$.

\item {\bf Case (b):} 

{\bf Transitions (i): $l \ne i$ or $M \ne j$:} In this case, $T_{i,j}(L, M) =
T_{i,j}(L, l)$. We know that $T_{i,j}(L, l)$ is the number of times that a
transition from state $i$ to state $j$ is used in a Viterbi path ending in
state $l$ at sequence position $L$. If we make a transition from state $l$ at
position $L$ to the \emph{End} state $M$ at sequence position $L$ and if $l
\ne i$ or $M \ne j$, we do not make a transition from state $i$ to state $j$
and thus do not need to modify the number of counts, i.e.\ $T_{i,j}(L, M) =
T_{i,j}(L, l)$. Also in case (a), we are now at the end of the input sequence
$X$ and the Viterbi path $\Pi^*(X)$ and thus have $T_{i,j}(L,M) = T_{i,j}^q(X, \Pi^*(X))$.
\end{itemize}

{\bf End of proof.}

As is clear from the above description of the algorithm, the calculation of
the $v_m$, $T_{i,j}$ and $E_i$ values for sequence position $k$ requires only
the respective values for the previous sequence position $k-1$, i.e.\ the
memory requirement can be linearized with respect to the sequence length.

For an HMM with $M$ states and a training sequence of length $L$ and for every
free parameter of the HMM that we want to train, we thus need in every
iteration $\mathcal{O}(M)$ memory to store the $v_m$ values and
$\mathcal{O}(M)$ memory to store the cumulative counts for the free parameter
itself, e.g.\ the $T_{i,j}$ values for a particular transition from state $i$
to state $j$. For an HMM, the memory requirement of the training using the new
algorithm is thus independent of the length of the training sequence.

For training one free parameter in the HMM with the above algorithm, each
iteration requires $\mathcal{O}(M T_{max} L)$ time to calculate the $v_m$
values and to calculate the cumulative counts.  If $Q$ is the total number of
free parameters in the model and if we choose $P$ of these parameters to be
trained in parallel, i.e.\ $P \in \{1, \ldots Q\}$ and $Q/P \in \mathbb{N}$,
the memory requirement increases slightly to $\mathcal{O}(M P)$ and the time
requirement becomes $\mathcal{O}(M T_{max} L \frac{Q}{P})$. This algorithm can
therefore be readily adjusted to trade memory and time requirements, e.g.\ to
maximize speed by using the maximum amount of available memory.

This can be directly compared to the default algorithm for Viterbi training
described above with first calculates the entire Viterbi matrix and which
requires $\mathcal{O}(ML)$ memory and $\mathcal{O}(T_{max} L M)$ time to
achieve the same.  Our new algorithm thus has the significant advantage of
linearizing the memory requirement with respect to the sequence length while
keeping the time requirement the same, see Table~1 for a detailed overview.
Our new algorithm is thus as memory efficient as Viterbi training using the
Hirschberg algorithm, while being more time efficient, significantly easier to
implement and applicable to all $n$-HMMs, including the case $n=1$.

\subsection*{A linear-memory algorithm for stochastic EM~training}

One alternative to Viterbi training is Baum-Welch training~\cite{baum1972},
which is an expectation maximization (EM) algorithm~\cite{dlr1977}. As Viterbi
training, Baum-Welch training is an iterative procedure. In each iteration of
Baum-Welch training, the estimated number of counts for each transition and
emission is derived by considering \emph{all} possible state paths for a given
training sequence in the model rather than only the single Viterbi path. As
discussed in the introduction, there already exists an efficient algorithm for
Baum-Welch training which linearizes the memory requirement with respect to
the sequence length and which is also relatively easy to implement.

One variant of Baum-Welch training is called stochastic
EM~algorithm~\cite{Bishop2006}. Unlike Viterbi training which considers only a
\emph{single} state path and unlike Baum-Welch training which considers
\emph{all} possible state paths for every training sequence, the stochastic
EM~algorithm derives new parameter values from a \emph{fixed} number of $K$
state paths (each of which is denoted $\Pi^s(X)$) that are sampled for each
training sequence from the posterior distribution $P(\Pi | X)$. Similar to
Viterbi and Baum-Welch training, the stochastic EM~algorithm employs an
iterative procedure. As for Baum-Welch training, the iterations are stopped
once a maximum number of iterations have been reached or once the change in
the log-likelihood is sufficiently small.

In strict analogy to the notation we introduced for Viterbi training,
$E^q_i(y, X, \Pi^s(X))$ denotes the number of times that state $i$ reads
symbol $y$ from input sequence $X$ in a sampled state path $\Pi^s(X)$ given
the HMM with parameters from the $q$-th iteration. Similarly, $T^q_{i,j}(X,
\Pi^s(X))$ denotes the number of times that a transition from state $i$ to
state $j$ is used in a sampled state path $\Pi^s(X)$ for sequence $X$ given
the HMM with parameters from the $q$-th iteration.


As usual, the superscript $q$ indicates from which iteration the underlying
parameters of the HMM derive. If we consider all $N$ sequences of the training
set $\mathcal{X} = \{X^1, \ldots X^N\}$ and sample $K$ state paths
$\Pi_k^s(X^n)$, $k \in \{1, \ldots K\}$, for each sequence $X^n$ in the
training set, the step which updates the values of the transition and emission
probabilities can be written as:

\begin{eqnarray}
  t^{q+1}_{i,j} & = & \frac{\sum_{n=1}^N \sum_{k=1}^K T^q_{i,j}(X^n, \Pi^s_k(X^n))}
                         {\sum_{j'=1}^M \sum_{n=1}^N \sum_{k=1}^K T^q_{i,j'}(X^n, \Pi^s_k(X^n))} \nonumber \\ 
  e^{q+1}_i(y) & = & \frac{\sum_{n=1}^N \sum_{k=1}^K E^q_i(y, X^n, \Pi^s_k(X^n))}
                         {\sum_{y' \in \mathcal{A}} \sum_{n=1}^N \sum_{k=1}^K E^q_i(y', X^n, \Pi^s_k(X^n))} \nonumber
\end{eqnarray}

These expressions are strictly analogous to equations~\ref{eq:T_update} and
\ref{eq:E_update} that we introduced for Viterbi training. As before, these
assume that we know the values of $T^q_{i,j}(X^n, \Pi^s_k(X^n))$ and $E^q_i(y,
X^n, \Pi^s_k(X^n))$, i.e.\ how often each transition and emission is used in
each sampled state path $\Pi^s_k(X^n)$ for every training sequence $X^n$.

\subsubsection*{Obtaining the counts from the  forward algorithm and stochastic back-tracing}

It is well-known that we can obtain the above counts $T_{i,j}(X, \Pi^s(X))$
and $E_i(y, X, \Pi^s(X))$ for a given training sequence $X$, iteration $q$ and
a sampled state path $\Pi^s(X)$ by using a combination of the forward
algorithm and stochastic back-tracing~\cite{Durbin1998, Bishop2006}. For this,
we first calculate all values in the two-dimensional forward matrix using the
forward algorithm and then invoke the stochastic back-tracing procedure to
sample a state-path $\Pi^s(X)$ from the posterior distribution $P(\Pi | X)$.

We will now explain these two algorithms in detail in order to facilitate the
introduction of our new algorithm. In the following,

\begin{itemize}
\item $f_i(k)$ denotes the sum of probabilities of all state paths that have
  read training sequence $X$ up to and including sequence position $k$ and
  that end in state $i$, i.e.\ $f_i(k) = P(x_1, \ldots, x_k, s(x_k) = i)$,
  where $s(x_k)$ denotes the state that reads sequence position $x_k$ from
  input sequence $X$. We call $f_i(k)$ the forward probability 
  for sequence position $k$ and state $i$.
\item $p_i(k, m)$ denotes the probability of selecting state $m$ as the
  previous state while being in state $i$ at sequence position $k$ (i.e.\
  sequence position $k$ has already been read by state $i$), i.e.\ $p_i(k, m)
  = P(\pi_{k-1} = m | \pi_k = i)$. For a given sequence position $k$ and state
  $i$, $p_i(k,m)$ defines a probability distribution over previous states as
  $\sum_m p_i(k,m) = 1$.
\end{itemize}

The forward matrix is calculated using the forward algorithm~\cite{Durbin1998}:

{\bf Initialization}: at the start of the input sequence, consider all states
$m \in \mathcal{S}$ in the model and set
\begin{eqnarray}
f_m(0) & = & 
\left\{
\begin{array}{rcl}
1 & m = 0 \nonumber\\
0 & m \neq 0
\end{array}
\right. \nonumber
\end{eqnarray}

{\bf Recursion}: loop over all positions $k$ from 1 to $L$ in the input
sequence and loop, for each such sequence position $k$, over all states $m \in
\mathcal{S} \backslash \{0\} = \{1, \ldots, M\}$ and set
\begin{eqnarray}
f_m(k) = e_m(x_{k}) \cdot \sum^M_{n=0}f_n(k-1)\cdot t_{n,m} \label{eq:recursion}
\end{eqnarray}

{\bf Termination}: at the end of the input sequence, i.e.\ for $k = L$ and $m
= M$ the \emph{End} state, set
\begin{eqnarray}
P(X) = f_{M}(L) = \sum^M_{n=0}f_n(L_x)\cdot t_{n,M} \nonumber
\end{eqnarray}

Once we have calculated all forward probabilities $f_i(k)$ in the
two-dimensional forward matrix, i.e.\ for all states $i$ in the model and all
positions $k$ in the given training sequence $X$, we can then use the
stochastic back-tracing procedure~\cite{Durbin1998} to sample a state path
from the posterior distribution $P(\Pi | X)$.

The stochastic back-tracing starts at the end of the input sequence, i.e.\ at
sequence position $k = L$, in the $\textsl{End}$ state, i.e.\ $i = M$, and
selects state $m$ as the previous state with probability:

\begin{eqnarray}
p_i(k,m) & = & 
\left\{
\begin{array}{ll}
\frac{f_m(k-1) \cdot e_i(x_k) \cdot t_{m,i}}{f_i(k)} & \textrm{if state } i \textrm{ is not silent} \\
\frac{f_m(k)   \cdot t_{m,i}}{f_i(k)}                & \textrm{if state } i \textrm{ is silent} 
\end{array}
\right. \label{eq:sampling}
\end{eqnarray}

This procedure is continued until we reach the start of the sequence and the
$Start$ state. The resulting succession of chosen previous states corresponds
to one state path $\Pi^s(X)$ that was sampled from the posterior distribution
$P(\Pi | X)$.

The denominator in equation~(\ref{eq:sampling}) corresponds to the sum of
probabilities of all state paths that finish in state $i$ at sequence position
$k$, whereas the nominator corresponds to the sum of probabilities of all
state paths that finish in state $i$ at sequence position $k$ \emph{and} that
have state $m$ as the previous state. 


When being in state $i$ at sequence position $k$, we can therefore use this
ratio to sample which previous state $m$ we should have come from.

As this stochastic back-tracing procedure requires the entire matrix of
forward values for all states and all sequence positions, the above algorithm
for sampling a state path requires $\mathcal{O}(ML)$ memory and $\mathcal{O}(M
T_{max} L)$ time in order to first calculate the matrix of forward values and
then $\mathcal{O}(L)$ memory and $\mathcal{O}(L T_{max})$ time for sampling a
single state path from the matrix. Note that additional state paths can be
sampled without having to recalculate the matrix of forward values. For
sampling $K$ state paths for the same sequence in a given iteration, we thus
need $\mathcal{O}((M + K) T_{max} L)$ time and $\mathcal{O}(ML)$ memory, if we
do not to store the sampled state paths themselves.

If our computer has enough memory to use the forward algorithm and the
stochastic back-tracing procedure described above, each iteration in the
training algorithm would require $\mathcal{O}(M \max_i\{L_i\} + K
\max_i\{L_i\})$ memory and $\mathcal{O}(M T_{max} \sum_{i=1}^N L_i + K
\sum_{i=1}^N L_i)$ time, where $\sum_{i=1}^N L_i$ is the sum of the $N$
sequence lengths in the training set $\mathcal{X}$ and $\max_i\{L_i\}$ the
length of the longest sequence in training set $\mathcal{X}$. As we do not
have to keep the $K$ sampled state paths in memory, the memory requirement can
be reduced to $\mathcal{O}(M \max_i\{L_i\})$.

For many bioinformatics applications, however, where the number of states in
the model $M$ is large, the connectivity $T_{max}$ of the model high or the
training sequences are long, these memory and time requirements are too large
to allow automatic parameter training using stochastic EM~training.

\subsubsection*{Obtaining the counts in a more efficient way}

Our previous observations (V1) to (V5) that led to the linear-memory algorithm
for Viterbi training can be replaced by similar observations for stochastic
EM~training:

\begin{itemize}
\item[(S1)] If we consider the description of the forward algorithm above, in
  particular the recursion in Equation~(\ref{eq:recursion}), we realize that
  the calculation of the forward values can be continued by retaining only the
  values for the previous sequence position.

\item [(S2)] If we have a close look at the description of the stochastic back-tracing
  algorithm, in particular the sampling step in Equation~(\ref{eq:sampling}), we
  observe that the sampling of a previous state only requires the forward
  values for the current and the previous sequence position. So, provided we
  are at a particular sequence position and in a particular state, we can sample the state at
  the previous sequence position, if we know all forward values for 
  the previous sequence position.

\item[(S3)] If we want to sample a state path $\Pi^s(X)$ from the posterior
  distribution $P(\Pi | X)$, we have to start at the \emph{end} of the
  sequence in the $\textsl{End}$ state, see the description above and
  Equation~(\ref{eq:sampling}) above. (The only valid alternative for sampling
  state paths from the posterior distribution would be to use the backward
  algorithm~\cite{Durbin1998} instead of the forward algorithm and to then
  start the stochastic back-tracing procedure at the \emph{start} of the
  sequence in the $Start$ state.)
\end{itemize}

Observations~(S1) and (S2) above imply that local information suffices to
continue the calculation of the forward values (S1) and to sample a previous
state (S2) if we already are in a particular state and sequence position,
whereas observation~(S3) reminds us that in order to sample from the correct
probability distribution, we have to \emph{start} the sampling at the
\emph{end} of the training sequence. Given these three observations, it is ---
as before for Viterbi training --- not obvious how we can come up with a
computationally more efficient algorithm. In order to realize that a more
efficient algorithm does exist, one also has to note that:

\begin{itemize}
\item[(S4)] While calculating the forward values in the memory-efficient way
  outlined in (S1) above, we can \emph{simultaneously} sample a
  previous state for every combination of a state and a sequence position that
  we encounter in the calculating of the forward values. This is possible
  because of observation (S2) above.

\item[(S5)] In every iteration $q$ of the training procedure, we only need to
  know the values of $T^q_{i,j}(X, \Pi^s(X))$ and $E^q_i(y, X, \Pi^s(X))$, i.e.\ how
  \emph{often} each transition and emission appears in each sampled state
  path $\Pi^s(X)$ for every training sequence $X$, but not \emph{where} in the 
  matrix of forward values the transition or emission was used.
\end{itemize}

Given all observations~(S1) to (S5) above, we can now formally write down a
new algorithm which calculates $T^q_{i,j}(X, \Pi^s(X))$ and $E^q_i(y, X,
\Pi^s(X))$ in a computationally more efficient way. In order to simplify the
notation, we consider one particular training sequence $X = (x_1, \ldots x_L)$
of length $L$ and omit the superscript for the iteration $q$, as both remain the
same throughout the following algorithm. In the following, $T_{i,j}(k, m)$
denotes the number of times the transition from state $i$ to state $j$ is used
in a sampled state path that finishes at sequence position $k$ in state $m$
and $E_i(y, k, m)$ denotes the number of times state $i$ read symbol $y$ in a
sampled state path that finishes at sequence position $k$ in state $m$.  As
defined earlier, $f_i(k)$ denotes the forward probability for sequence
position $k$ and state $i$, $p_i(k,m)$ is the probability of selecting state
$m$ as the previous state while being in state $i$ at sequence position $k$,
$i,j,n \in \mathcal{S}$ and $y \in \mathcal{A}$.

{\bf Initialization}: at the start of the training sequence $X$ and for 
all states $m \in \mathcal{S}$, set
\begin{eqnarray}
f_m(0) & = & 
\left\{
\begin{array}{rcl}
1 & m = 0  \nonumber\\
0 & m \neq 0
\end{array}
\right. \nonumber \\
T_{i,j} (0, m) & = & 0 \nonumber \\
E_i(y, 0, m) &=& 0 \nonumber
\end{eqnarray}

{\bf Recursion}: loop over all positions $k$ from 1 to $L$ in the training
sequence $X$ and loop, for each such sequence position $k$, 
over all states $m \in \mathcal{S} \backslash \{0\} = \{1, \ldots, M\}$ and set
\begin{eqnarray}
f_m(k)          & = & e_m(x_{k}) \cdot \sum^M_{n=0}f_n(k-1)\cdot t_{n,m}        \nonumber \\
p_m(k, n)       & = & \frac{e_m(x_{k}) \cdot f_n(k-1) \cdot t_{n,m}}{f_m(k)}  \nonumber \\
T_{i,j}(k, m)    & = & T_{i,j}(k-1, l) + \delta_{l,i} \cdot \delta_{m,j}          \nonumber \\
E_i(y, k, m)    & = & E_i(y, k-1, l) + \delta_{m,i} \cdot \delta_{y, x_k}          \nonumber
\end{eqnarray}

where $l$ denotes the state at previous sequence position $k-1$ that was
sampled from the probability distribution $p_m(k, n)$, $n \in \mathcal{S}$,
while being in state $m$ at sequence position $k$.

{\bf Termination}: at the end of the input sequence, i.e.\ for $k=L$ and $m =
M$ the \emph{End} state, set
\begin{eqnarray}
f_{M}(L)      & = & \sum^M_{n=0}f_n(L_x)\cdot t_{n,M}        \nonumber \\
p_{M}(L, n)   & = & \frac{f_n(L) \cdot t_{n, M}}{f_{M}(L)} \nonumber \\
T_{i,j}(L, M) & = & T_{i,j}(L, l) + \delta_{l,i} \cdot \delta_{M,j}             \nonumber \\
E_i(y, L, M) & = & E_i(y, L, l)          \nonumber
\end{eqnarray}

where $l$ now denotes the state at sequence position $L$ that was sampled
from the probability distribution $p_M(L, n)$, $n \in \mathcal{S}$, while
being in the $\textsl{End}$ state $M$ at sequence position $L$, i.e.\ at the
end of the training sequence.

The above algorithm yields, $T_{i,j}(L,M) = T_{i,j}^q(X, \Pi^s(X))$ and
$E_i(y, L, M) = E_i^q(y, X, \Pi^s(X))$ (and $f_M(L) = P^q(X)$), i.e.\ we know
how often a transition from state $i$ to state $j$ was used and how often
symbol $y$ was read by state $i$ in a state path $\Pi^S(X)$ sampled from the
posterior distribution $P(X| \Pi)$ in iteration $q$ for sequence $X$.

{\bf Theorem 2:} The above algorithm yields $T_{i,j}(L,M) = T_{i,j}^q(X,
\Pi^s(X))$ and $E_i(y, L, M) = E_i^q(y, X, \Pi^s(X))$.

{\bf Proof:} The proof for this theorem is very similar to the proof of
theorem~1 for Viterbi training and therefore omitted. The key differences are,
first, that $l$ here corresponds to the state at the previous sequence
position that is \emph{sampled from a probability distribution} rather than
deterministically determined and, second, that $\Pi^s$ here corresponds to a
sampled state path rather than a deterministically derived Viterbi path
$\Pi^*$. 


{\bf End of proof.}

As is clear from the above algorithm, the calculation of the $f_m$, $p_m$, $T_{i,j}$
and $E_i$ values for sequence position $k$ requires only the respective values
for the previous sequence position $k-1$, i.e.\ the memory requirement can be
linearized with respect to the sequence length.

For an HMM with $M$ states, a training sequence of length $L$ and for every
free parameter to be trained, we thus need $\mathcal{O}(M)$ memory to store
the $f_m$ values, $\mathcal{O}(T_{max})$ memory to store the $p_m$ values
and $\mathcal{O}(M)$ memory to store the cumulative counts for the free
parameter itself in every iteration, e.g.\ the $T_{i,j}$ values for a
particular transition from state $i$ to state $j$. If we sample $K$ state
paths, we have to store the cumulative counts from different state paths
\emph{separately}, i.e.\ we need $K$ times more memory to store the cumulative
counts for each free parameter, but the memory for storing the $f_m$ and the
$p_m$ values remains the same. Overall, if $K$ state paths are being sampled
in each iteration, we thus need $\mathcal{O}(M)$ memory to store the $f_m$
values, $\mathcal{O}(T_{max})$ memory to store the $p_m$ values and
$\mathcal{O}(M K)$ memory to store the cumulative counts for the free
parameter itself in every iteration. For an HMM, the memory requirement of the
new training algorithm is thus independent of the length of the training
sequence.

For training one free parameter in the HMM with the above algorithm, each
iterations requires $\mathcal{O}(M T_{max} L)$ time to calculate the $f_m$ and
the $p_m$ values and to calculate the cumulative counts for one training
sequence. If $K$ state paths are being sampled in each iteration, the time
required to calculate the cumulative counts increases to $\mathcal{O}(M
T_{max} L K)$, but the time requirements for calculating the $f_m$ and $p_m$
values remains the same.

For sampling $K$ state paths for the same input sequence and training one free
parameter, we thus need $\mathcal{O}(M K + T_{max})$ memory and $\mathcal{O}(M
T_{max} L K)$ time for every iteration. If the model has $Q$ parameters and if
$P$ of these parameters are to be trained in parallel, i.e.\ $P \in \{1,
\ldots Q\}$ and $Q/P \in \mathbb{N}$, the memory requirement increases
slightly to $\mathcal{O}(M K P + T_{max})$ and the time requirement becomes
$\mathcal{O}(M T_{max} L K \frac{Q}{P})$.  As for Viterbi training, the
linear-memory algorithm for stochastic EM~training can therefore be readily
used to trade memory and time requirements, e.g.\ to maximize speed by using
the maximum amount of available memory, see Table~1 for a detailed overview.

This can be directly compared to the algorithm described in 2.1 with requires
$\mathcal{O}(ML)$ memory and $\mathcal{O}(T_{max} L (M + K))$ time to do the
same.  Our new algorithm thus has the significant advantage of linearizing the
memory requirement and making it independent of the sequence length for HMMs
while increasing the time requirement only by a factor of $\frac{M K}{M + K}$,
i.e.\ decreasing it when only one state path $K = 1$ is sampled.

\subsection*{Examples}


The algorithms that we introduce here can be used to train any HMM.  The
previous sections discuss the theoretical properties of the different
parameter training methods in detail which are summarized in Table~1.

Even though the theoretical properties of the respective algorithms are
independent of any particular HMM, the outcome of the different types of
parameter training in terms of prediction accuracy and parameter convergence
may very well depend on the features of a particular HMM. This is because the
quantities that can be shown to be (locally) optimized by some training
algorithms do not necessarily translate into an optimized prediction accuracy
as defined by us here.

In order to investigate how well the different methods do in practice in terms
of prediction accuracy and parameter convergence, we implemented Viterbi
training, Baum-Welch training and stochastic EM~training for three small
example HMMs. For each model, we implemented the linear-memory algorithm for
Baum-Welch training published earlier as well as the linear-memory algorithms
for Viterbi training and stochastic EM~training presented here.

In the first step, we use each model with the original parameter values to
generate the sequences of the data set. We then randomly choose initial
parameter values to initialize the HMM for parameter training. Each type of
parameter training is performed three times using $2/3$ of the un-annotated
data set as training set and the remaining $1/3$ of the data set for
performance evaluation, i.e.\ we perform three cross-evaluation experiments
for each model.

\paragraph*{Example 1: The dishonest casino}

As first case, we consider the well-known example of the dishonest
casino~\cite{Durbin1998}, see Figure~\ref{fig:dishonest_casino}. This casino
consists of a fair (state $\textrm{F}$) and a loaded dice (state
$\textrm{L}$). The fair dice generates numbers from $\mathcal{A} =
\{1,2,3,4,5,6\}$ with equal probability, whereas the loaded dice generates the
same numbers in a biased way.  The properties of the dishonest casino are
readily captured in a four-state HMM with 8~transition and 12~emission
probabilities, six each for each non-silent state $F$ and $L$. Parameterizing
the emission and transition probabilities of this HMM results in two
independent transition probabilities and 10~independent emission
probabilities, i.e.\ altogether 12~values to be trained. In order to avoid
premature termination of parameter training, we use pseudo-counts of~1 for
every parameter to be trained.

The data set for this model consists of 300~sequences of 5000~bp length
each. The results of the training experiments are shown in
Figures~\ref{fig:dishonest_performance} and \ref{fig:dishonest_convergence}.

\paragraph*{Example 2: The extended dishonest casino}

In order to investigate a HMM with a more complicated regular grammar, we
extended the above example of the dishonest casino so it can now use the
loaded dice (state $\textrm{L}$) only in multiples of two and the fair dice
(state $\textrm{F}$) only in multiples of three, see
Figure~\ref{fig:extended_dishonest_casino}.

This extended HMM has seven states, the silent $\textrm{Start}$ and
$\textrm{End}$ states, two $\textrm{F}$ states and three $\textrm{L}$ states,
11~transition probabilities and 30~emission probabilities. Parameterizing the
HMM's probabilities yields two independent transition probabilities and
10~independent emission probabilities to be trained, i.e.\ 12~parameter
values. In order to avoid premature termination of parameter training, we use
pseudo-counts of~1 for every parameter to be trained.

The data set for this model consists of 300~sequences of 5000~bp length
each. The results for this extended model are shown in
Figures~\ref{fig:extended_dishonest_performance} and
\ref{fig:extended_dishonest_convergence}.


\paragraph*{Example 3: The CpG~island model}

In order to study the features for the different training algorithms for a
bioinformatics application, we also investigate an HMM that can be used to
detect CpG~islands in sequences of genomic~DNA~\cite{Durbin1998}, see
Figure~\ref{fig:CpG_island}. The model consists of 10~states, the silent
$\textrm{Start}$ and $\textrm{End}$ states, four non-silent states to model
regions inside CpG~islands (states $\textrm{A}^+$, $\textrm{C}^+$,
$\textrm{G}^+$ and $\textrm{T}^+$) and four non-silent states to model regions
outside CpG~islands (states $\textrm{A}^-$, $\textrm{C}^-$, $\textrm{G}^-$ and
$\textrm{T}^-$). The emission probabilities for each of the eight non-silent
states is a delta-function so that any particular state (say $\textrm{A}^+$ or
$\textrm{A}^-$) has an emission probability of $1$ for reading the
corresponding DNA~nucleotide (in this case $\tt{A}$) and a probability of zero
for all other nucleotides, i.e.\ $e_{X^+}(Y) = e_{X^-}(Y) = \delta_{X,Y}$ for
$X,Y \in \{{\tt A}, {\tt C}, {\tt G}, {\tt T}\}$. This implies that none of
the emission probabilities of this model thus requires training. With a total
of 80~transition probabilities the model is, however, highly connected as any
non-silent state is connected in both directions to any other non-silent
state. Parameterizing these transition probabilities results in 33~parameters,
32 of which were determined in training (the transition probability to go to
the $\textrm{End}$ state was fixed). In order to avoid premature termination
of parameter training, we use pseudo-counts of~1 for every parameter to be
trained.

The data set for this model consists of 180~sequences of 5000~bp length each.
Figures~\ref{fig:CpG_performance} and \ref{fig:CpG_convergence} show the
resulting performance.

\subsubsection*{Prediction accuracy and parameter convergence}

Our primary goal is to investigate how the prediction accuracy of the
different training algorithms varies as function of the number of iterations.
The prediction accuracy or performance is defined as the product of the
sensitivity and specificity. Figures~\ref{fig:dishonest_performance},
\ref{fig:extended_dishonest_performance} and \ref{fig:CpG_performance} show
the prediction accuracy as function of the number of iterations for all three
training methods for the respective model. 

Another important goal of parameter training is to recover the original
parameter values of the corresponding model.  We therefore also investigate
how well the trained parameter values converge to the original parameter
values, see Figures~\ref{fig:dishonest_convergence},
\ref{fig:extended_dishonest_convergence} and \ref{fig:CpG_convergence} show
the average difference between the trained and known parameter values as
function of the number of iterations for each training algorithm and the
respective model. Every data point is calculated by first determining the
average value of the absolute differences between the trained and known value
of each emission parameter (left figures) or transition parameter (right
figures) and then taking the average over the three experiments from the
three-fold cross-evaluation.

For the dishonest casino and the extended dishonest casino, stochastic
EM~training performs best, both in terms of performance and parameter
convergence. It is interesting to note that the results for sampling one,
three or five state paths per training sequence and per iteration are
essentially the same within error bars.  For these two models, Viterbi
training converges fastest, i.e.\ the Viterbi paths remain the same from one
iteration to the next, but the point of convergence is sub-optimal in terms of
performance and in particular in terms of parameter convergence.  Baum-Welch
training does better than Viterbi training for these two models, but not as
well as stochastic BM~training as it requires more iterations to reach a lower
prediction accuracy and worse parameter convergence and as it exhibits the
largest variation with respect to the three cross-evaluation experiments. 
The latter is due to many high-scoring, sub-optimal state paths.  For the
CpG~island model, all training algorithms do almost equally well, with Viterbi
training converging fastest.

Table~2 summarizes the CPU time per iteration for the different training
algorithms and models. For all three models, stochastic EM~training is faster
than Baum-Welch training for one, three or five sampled state paths per
training sequence. Viterbi training is even a bit more time efficient than
stochastic EM~training when sampling one state path per training sequence.

Based on the results from these three small example models, we would thus
recommend using stochastic EM~training for parameter training.

\section*{Conclusion and discussion}

A wide range of bioinformatics applications are based on hidden Markov models.
Having computationally efficient algorithms for training the free parameters
of these models is key to optimizing the performance of these models and to
adapting the models to new data sets, e.g.\ biological data sets from a
different organism.

We here introduce two new algorithms which render the memory requirements for
Viterbi training and stochastic EM~training independent of the sequence
length. This is achieved by replacing the usual bi-directional two-step
procedure (which involves first calculating the Viterbi matrix and then
retrieving the Viterbi path (in case of Viterbi training) or first calculating
the forward matrix and the backward matrix before estimating counts (in case
of Baum-Welch training)) by a one-step procedure which scans each training
sequence only in a one-directional way. For an HMM with $M$ states and a
connectivity of $T_{max}$, a training sequence of length $L$ and one
iteration, our new algorithm reduces the memory requirement of Viterbi
training from $\mathcal{O}(M L)$ to $\mathcal{O}(M)$ while keeping the time
requirement of $\mathcal{O}(M T_{max} L)$ unchanged, see Table~1 for details.
For stochastic EM~training where $K$ is the number of state paths sampled for
every training sequence in every iteration, the memory requirements are (as,
typically, $L \gg K + 1 \ge K + T_{max}/M$) reduced from $\mathcal{O}(M L)$ to
$\mathcal{O}(M K + T_{max})$ while the time requirement per iteration
changes from $\mathcal{O}(T_{max} L (M + K))$ to $\mathcal{O}(T_{max} L M K)$
depending on the user-chosen value of $K$.  An added advantage of our two new
algorithms is they are easier to implement than the corresponding default
algorithms for Viterbi training and stochastic EM~training. In addition to
introducing the two new algorithms for Viterbi training and stochastic
EM~training, we also examine their practical merits for three small example
models by comparing them to the linear-memory algorithm for Baum-Welch
training which was introduced earlier. Based on our results from these three
(non-representative) models, we would recommend using stochastic EM~training
for parameter training.

We have implemented the new algorithms for Viterbi training and stochastic
EM~training as well as the linear-memory algorithm for Baum-Welch training
into our HMM-compiler {\sc HMMConverter}~\cite{Lam2009} which can be used to
set up a variety of HMM-based applications and which is freely available under
the GNU General Public License version 3 (GPLv3). Please see 
{\tt  http://people.cs.ubc.ca/\~{}$\!$irmtraud/training} for more information
and the source code.

We hope that the new parameter training algorithms introduced here will make
parameter training for HMM-based applications easier, in particular those in
bioinformatics.

\section*{Competing interests}

The authors declare that they have no competing interests.

\section*{Authors' contributions}

T.Y.L.\ and I.M.M.\ devised the new algorithms, T.Y.L.\ implemented them,
T.Y.L.\ and I.M.M.\ conducted the experiments, evaluated the experiments and
wrote the manuscript. All authors read and approved the final manuscript.

\section*{Acknowledgments}

\ifthenelse{\boolean{publ}}{\small}{Both authors would like to thank the
  anonymous referees for providing useful comments. We would also like to
  thank Anne Condon for giving us helpful feedback on our manuscript.  Both
  authors gratefully acknowledge support by a Discovery Grant of the Natural
  Sciences and Engineering Research Council, Canada, and by a Leaders
  Opportunity Fund of the Canada Foundation for Innovation to I.M.M.}


{\ifthenelse{\boolean{publ}}{\footnotesize}{\small}
\bibliographystyle{bmc_article}
\bibliography{document}}

\ifthenelse{\boolean{publ}}{\end{multicols}}{}


\clearpage



  \begin{figure}[p]
    \begin{center}
     \pgfimage[height=3cm]{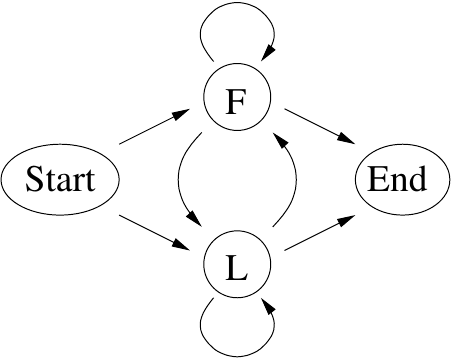}
     \end{center}
     \caption{\label{fig:dishonest_casino} {\bf HMM of the dishonest casino.}
       Symbolic representation of the HMM of the dishonest casino.  States are
       shown as circles, transitions are shown as directed arrows.  Please
       refer to the text for more details.}
  \end{figure}


  \begin{figure}[p]
    \begin{center}
     \pgfimage[height=6cm]{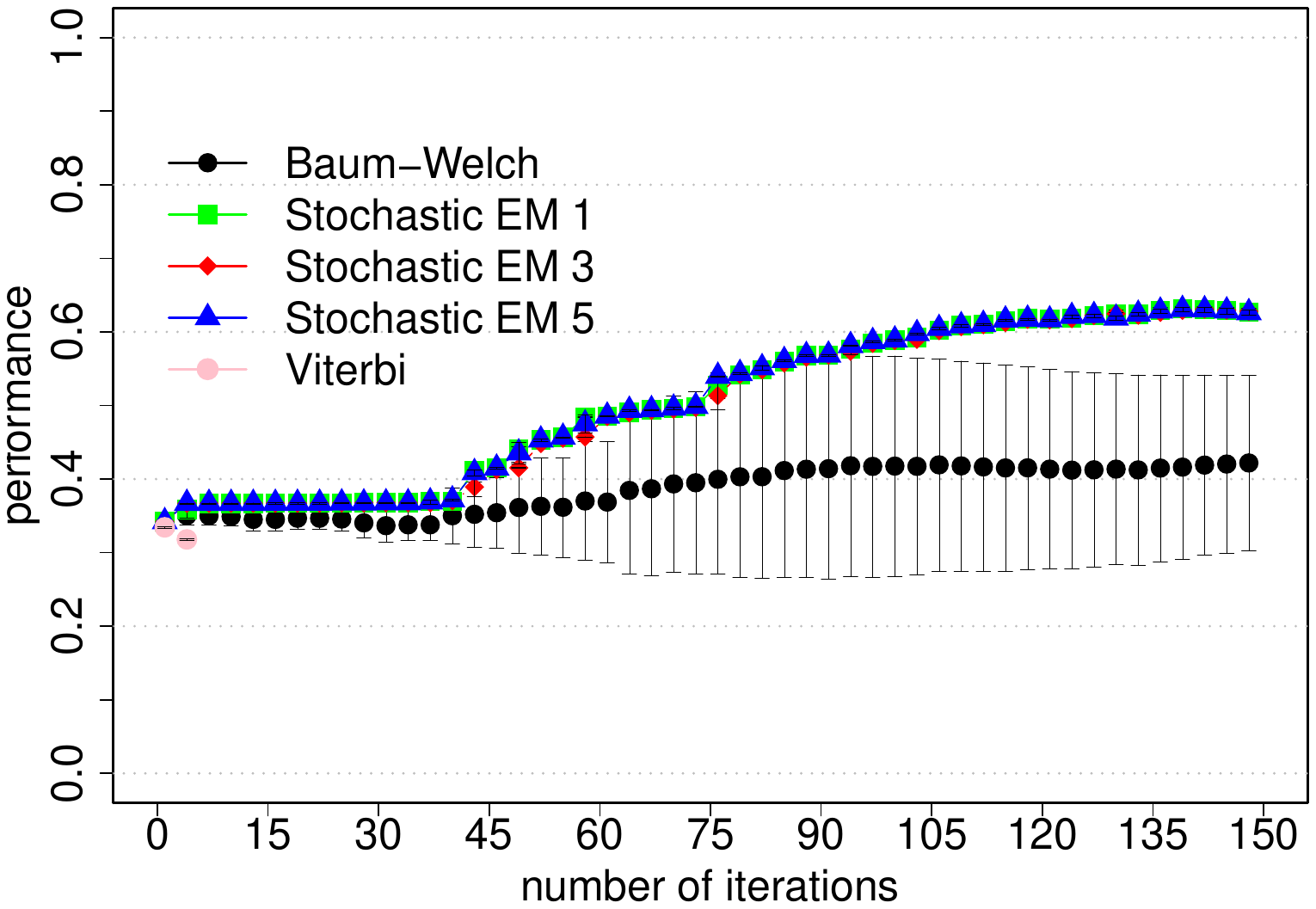}
     \end{center}
     \caption{\label{fig:dishonest_performance} {\bf Performance for the
         dishonest casino.} The average performance as function of the number
       of iterations for each training algorithm.  The performance is defined
       as the product of the sensitivity and specificity and the average is
       the average of three cross-evaluation experiments. For stochastic
       EM~training, a fixed number of state paths were sampled for each
       training sequence in each iteration (stochastic EM 1: one sampled state
       path, stochastic EM 3: three sampled state paths, stochastic EM 5: five
       sampled state paths). The error bars correspond to the standard
       deviation of the performance from the three cross-evaluation
       experiments. Please refer to the text for more information.}
  \end{figure}


 \begin{figure}[p]
    \begin{center}
    \pgfimage[height=6cm]{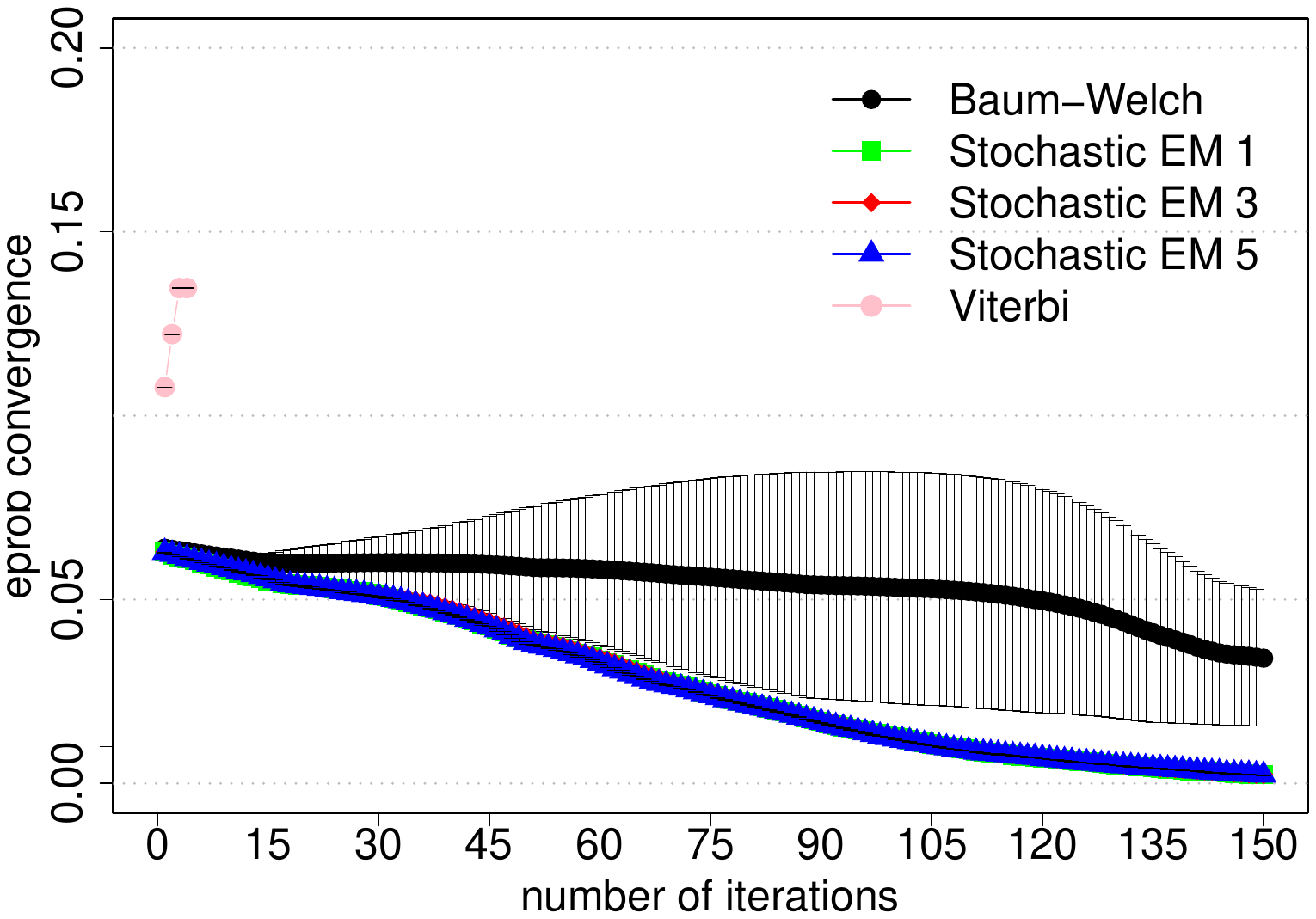}
    \pgfimage[height=6cm]{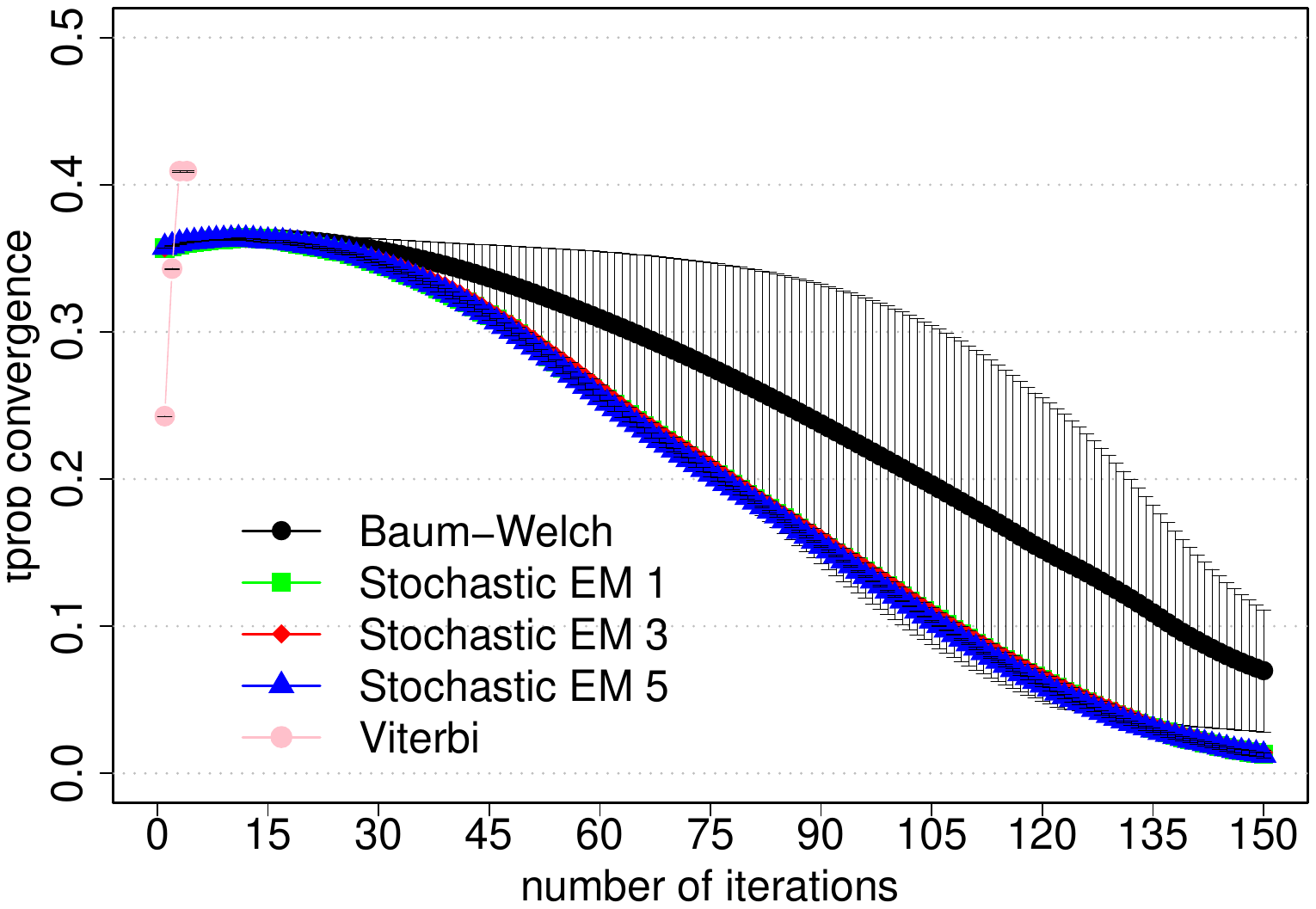}
  \end{center}
 \caption{\label{fig:dishonest_convergence} {\bf Parameter convergence for the
     dishonest casino.}   Average difference of the trained and known parameter values as function of
  the number of iterations for each training algorithm. For a given number of
  iterations, we first calculate the average value of the absolute differences
  between the trained and known value of each emission parameter (left figure)
  or transition parameter (right figure) and then take the average over the
  three experiments from the three-fold cross-evaluation. The error bars
  correspond to the standard deviation from the three cross-evaluation
  experiments. The algorithms have the same meaning as in Figure~2. 
  Please refer to the text for more information.}
 \end{figure}


  \begin{figure}[p]
    \begin{center}
     \pgfimage[height=3cm]{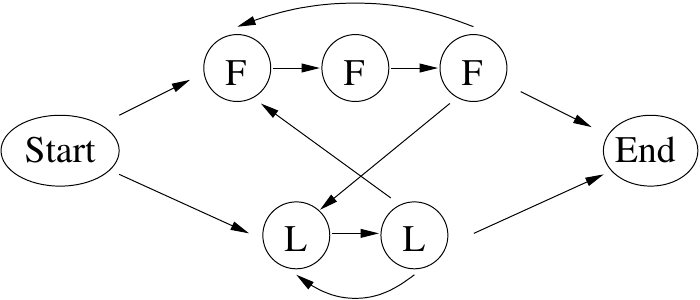}
    \end{center}
     \caption{\label{fig:extended_dishonest_casino} {\bf HMM of the extended
         dishonest casino.} Symbolic representation of the HMM of the extended
       dishonest casino.  States are shown as circles, transitions are shown
       as directed arrows.  Please refer to the text for more details.}
  \end{figure}


 \begin{figure}[p]
    \begin{center}
    \pgfimage[height=6cm]{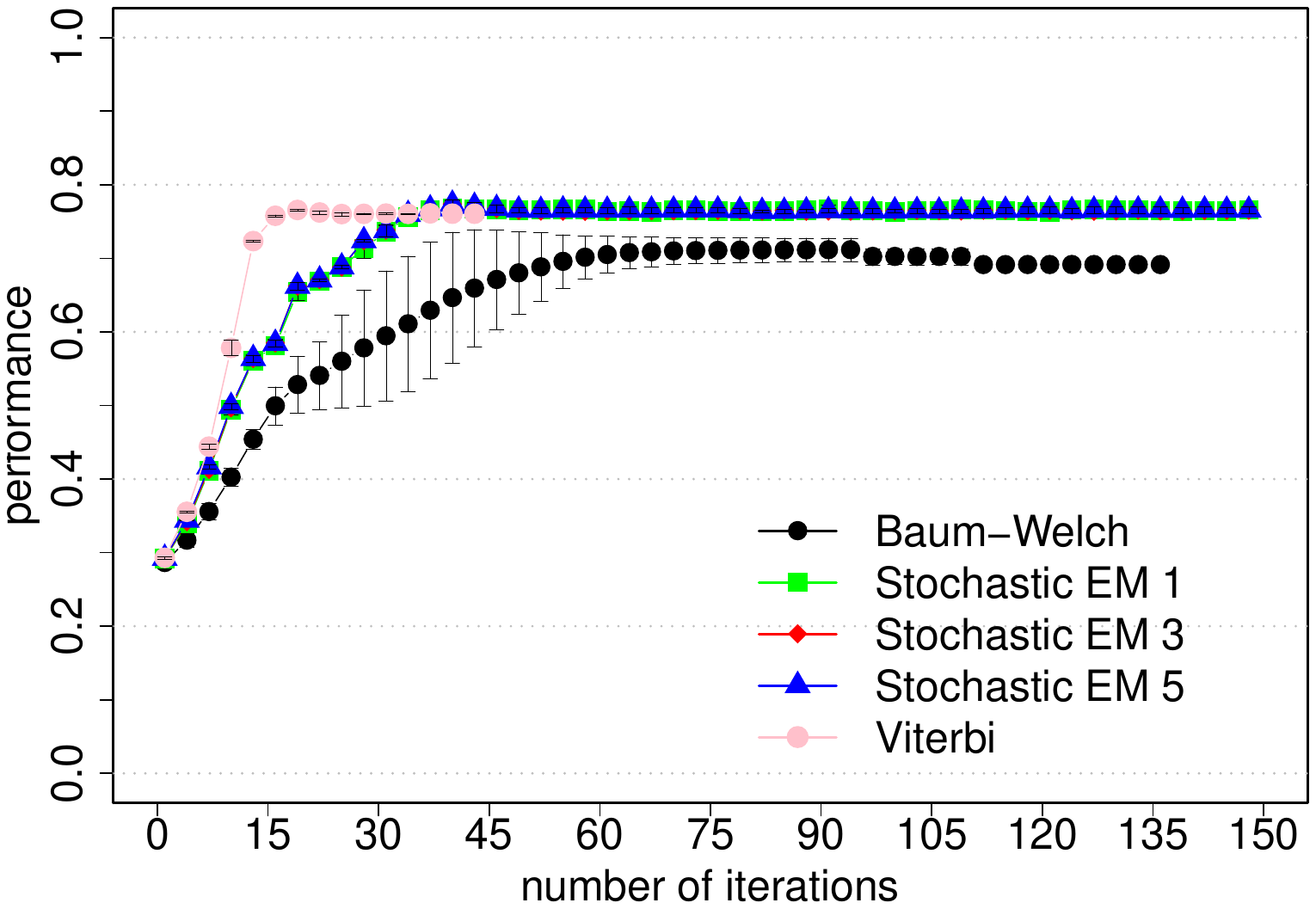}
    \end{center}
    \caption{\label{fig:extended_dishonest_performance} {\bf Performance for
        the extended dishonest casino.}  The average performance as function
      of the number of iterations for each training algorithm. The performance
      is defined as the product of the sensitivity and specificity and the
      average is the average of three cross-evaluation experiments. For
      stochastic EM~training, a fixed number of state paths were sampled for
      each training sequence in each iteration (stochastic EM 1: one sampled
      state path, stochastic EM 3: three sampled state paths, stochastic EM 5:
      five sampled state paths). The error bars correspond to the standard
      deviation of the performance from the three cross-evaluation
      experiments. Please refer to the text for more information.}
 \end{figure}


 \begin{figure}[p]
    \begin{center}
    \pgfimage[height=6cm]{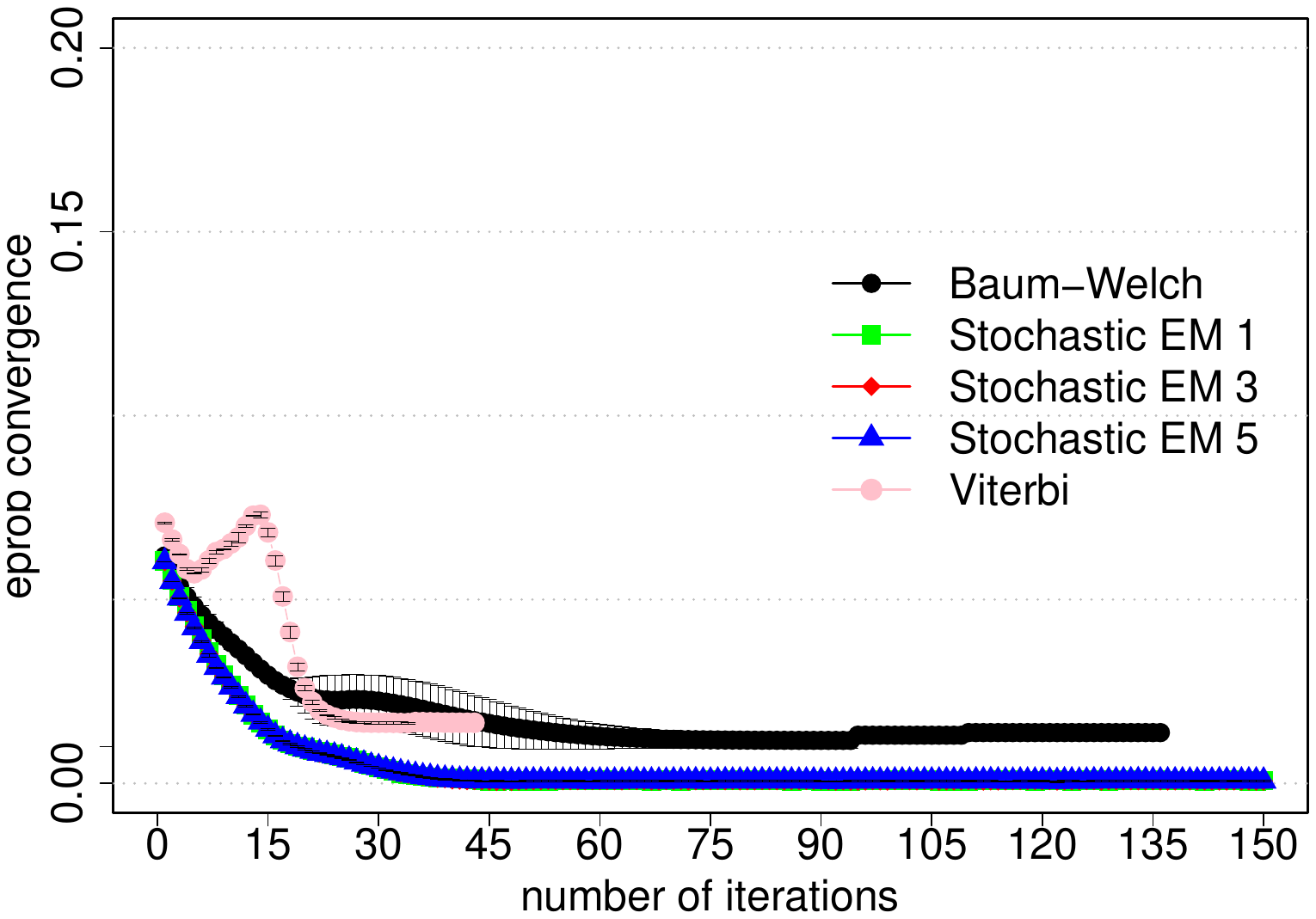}
    \pgfimage[height=6cm]{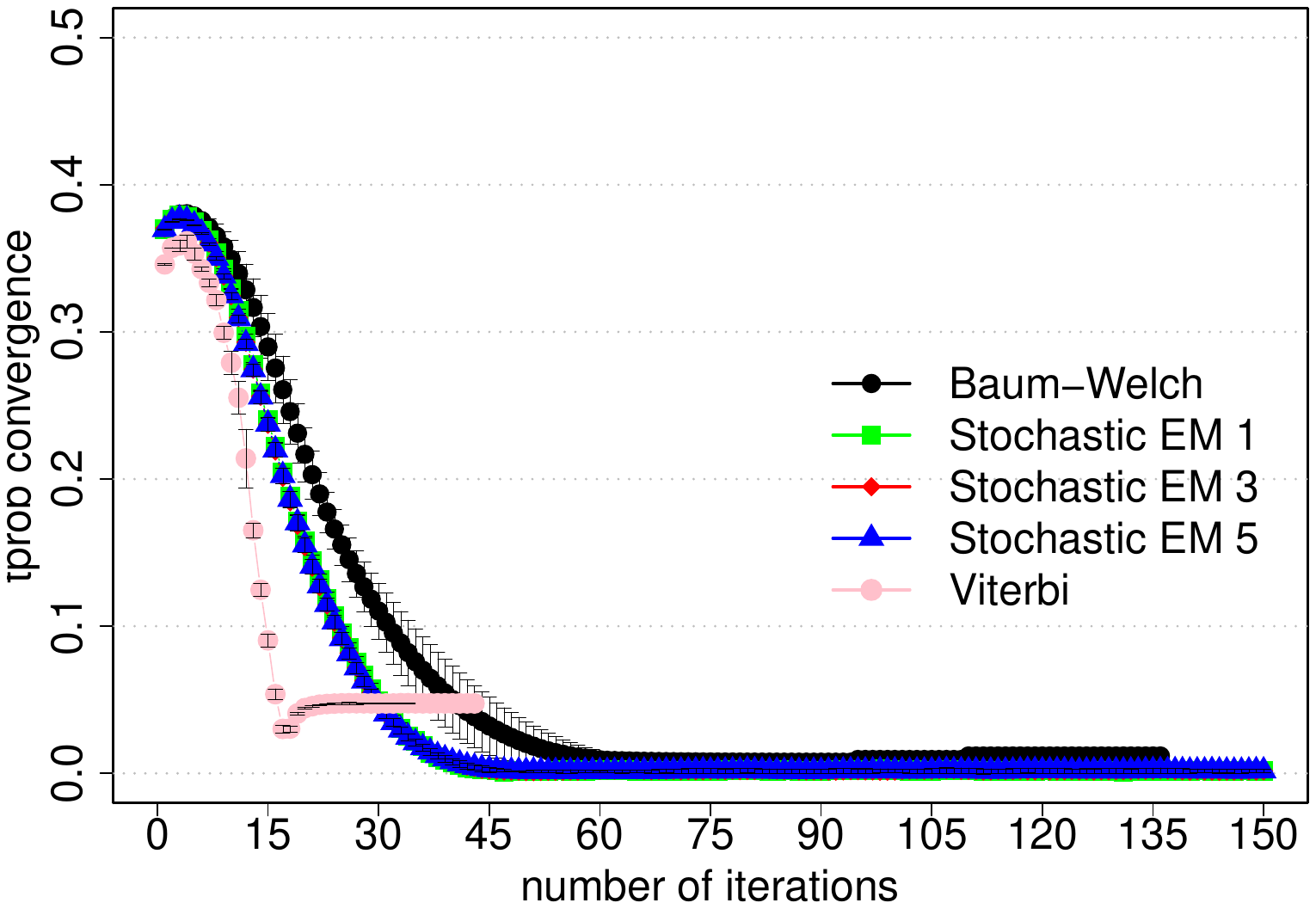}
    \end{center}
    \caption{\label{fig:extended_dishonest_convergence} {\bf Parameter
        convergence for the extended dishonest casino.}  Average difference of
      the trained and known parameter values as function of the number of
      iterations for each training algorithm. For a given number of
      iterations, we first calculate the average value of the absolute
      differences between the trained and known value of each emission
      parameter (left figure) or transition parameter (right figure) and then
      take the average over the three cross-evaluation experiments. The error
      bars correspond to the standard deviation from the three
      cross-evaluation experiments. The algorithms have the same meaning as in
      Figure~5.  Please refer to the text for more information.}  
 \end{figure}


  \begin{figure}[p]
    \begin{center}
     \pgfimage[height=4cm]{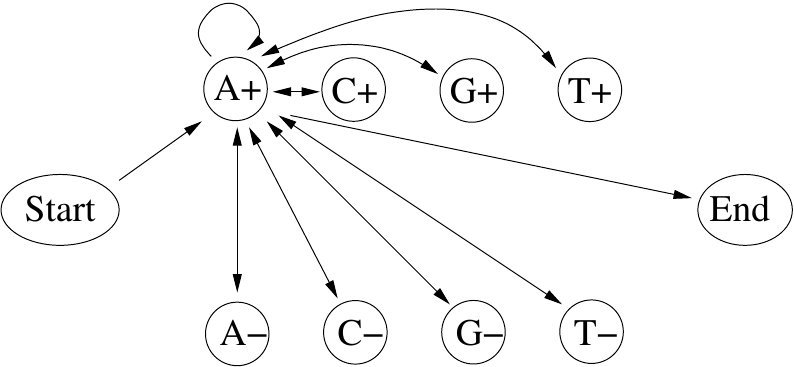}
    \end{center}
     \caption{\label{fig:CpG_island} {\bf CpG~island HMM.}  Symbolic
       representation of the CpG~island HMM.  States are shown as circles,
       transitions are shown as directed arrows. Every non-silent state can be
       reached from the $\textsl{Start}$ state and has a transition to the
       $\textsl{End}$ state. In addition, every non-silent state is connected
       in both directions to all non-silent states. For clarity, we here only
       show the transitions from the perspective of the $\textrm{A}^+$ state.
       Please refer to the text for more details.  }
  \end{figure}


  \begin{figure}[p]
    \begin{center}
     \pgfimage[height=6cm]{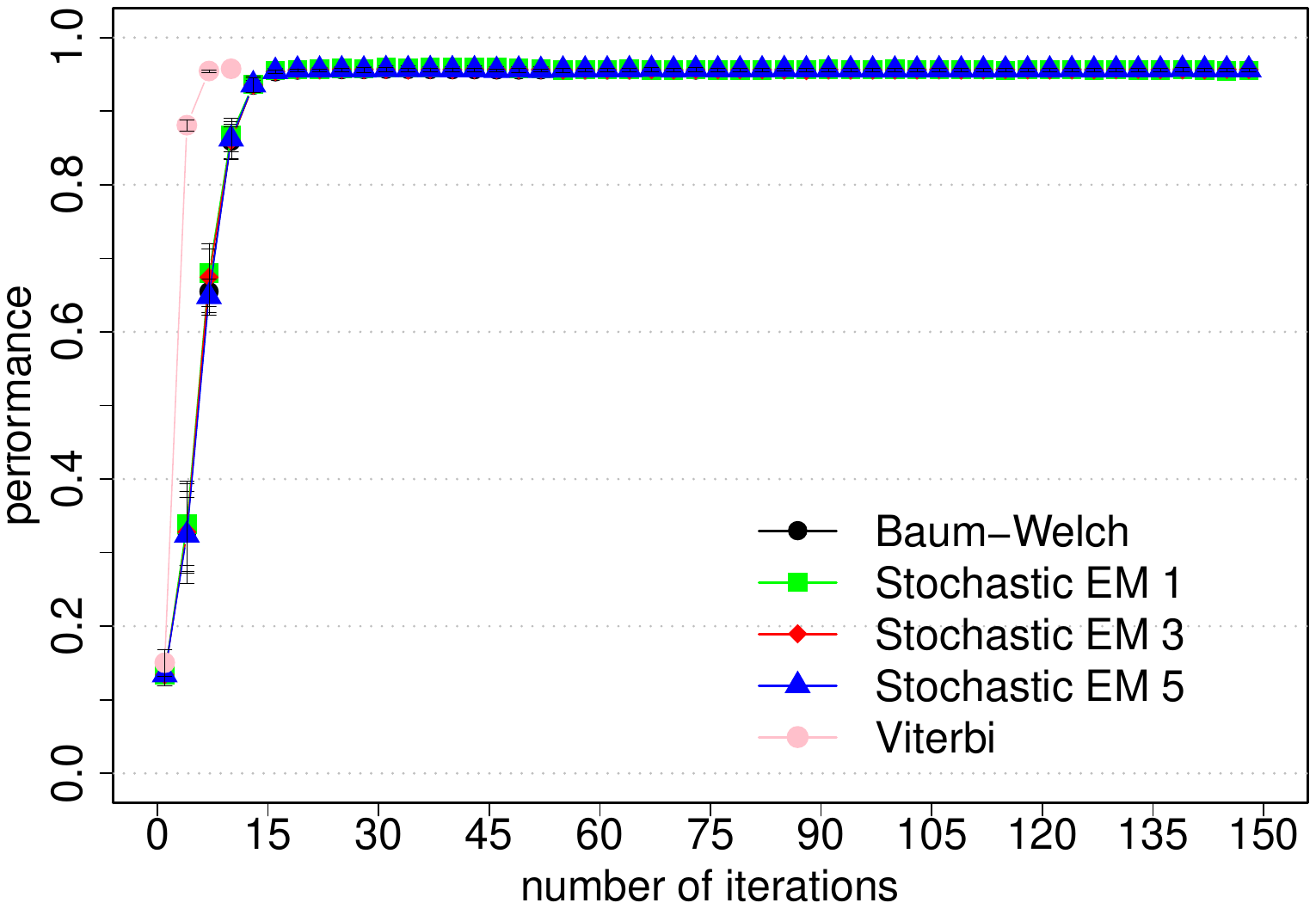}
    \end{center}
     \caption{\label{fig:CpG_performance} {\bf Performance for the CpG~island
         model.}  The average performance as function of the number of
       iterations for each training algorithm. The performance is defined as
       the product of the sensitivity and specificity and the average is the
       average of three cross-evaluation experiments. For stochastic
       EM~training, a fixed number of state paths were sampled for each
       training sequence in each iteration (stochastic EM 1: one sampled state
       path, stochastic EM 3: three sampled state paths, stochastic EM 5: five
       sampled state paths). The error bars correspond to the standard
       deviation of the performance from the three cross-evaluation
       experiments. Please refer to the text for more information.  }
  \end{figure}


 \begin{figure}[p]
    \begin{center}
    \pgfimage[height=6cm]{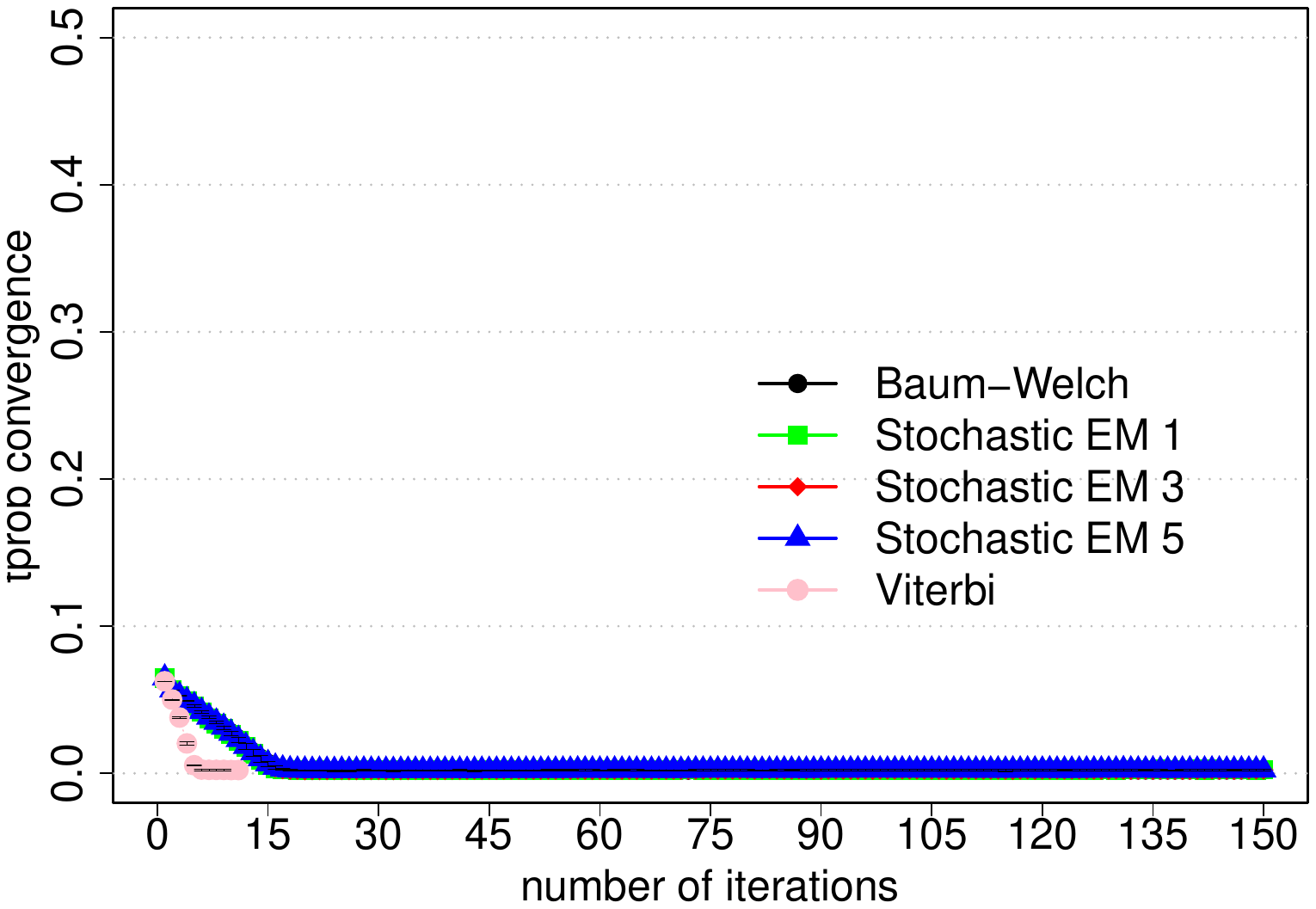}
    \end{center}
    \caption{\label{fig:CpG_convergence} {\bf Parameter convergence for the
        CpG~island model.}  Average difference of the trained and known
      parameter values as function of the number of iterations for each
      training algorithm. For a given number of iterations, we first calculate
      the average value of the absolute differences between the trained and
      known value of each transition parameter (this model does not have any
      emission parameters that require training) and then take the average
      over the three cross-evaluation experiments.  The error bars correspond
      to the standard deviation from the three cross-evaluation
      experiments. The algorithms have the same meaning as in Figure~8.
      Please refer to the text for more information.  }
 \end{figure}

\clearpage



   \begin{table}
     \begin{center}
       \begin{tabular}{l|llll}
         \multicolumn{5}{c}{training one parameter at a time} \\ \hline
type of training   &  algorithm              & time                            & memory                 & reference \\ \hline 
Viterbi            & Viterbi                 & $\mathcal{O}(T_{max} L M)$       & $\mathcal{O}(M L)$     & \cite{viterbi1967} \\
                   & Lam-Meyer               & $\mathcal{O}(T_{max} L M)$       & $\mathcal{O}(M)$   &  this paper \\[0.5em]  \hline
Baum-Welch         & Baum-Welch              & $\mathcal{O}(T_{max} L M)$       & $\mathcal{O}(M L)$        & \cite{Durbin1998} \\
                   & checkpointing           & $\mathcal{O}(T_{max} L M \log(L))$ & $\mathcal{O}(M \log(L))$ &\cite{Grice1997} \\ 
                   & linear-memory           & $\mathcal{O}(T_{max} L M)$       & $\mathcal{O}(M)$       & \cite{Miklos2005} \\ \hline

stochastic~EM      & forward \& back-tracing  & $\mathcal{O}(T_{max} L (M + K))$ & $\mathcal{O}(M L)$ & \cite{Bishop2006} \\
                   & Lam-Meyer               & $\mathcal{O}(T_{max} L M K)$     & $\mathcal{O}(M K + T_{max})$ & this paper \\[1.5em] 
\multicolumn{5}{c}{training $P$ of $Q$ parameters at the same time with $P \in \{1, \ldots, Q\}$ and $Q/P \in \mathbb{N}$} \\ \hline
Viterbi            & Viterbi                 & $\mathcal{O}(T_{max} L M Q/P)$       & $\mathcal{O}(M L)$     & \cite{viterbi1967} \\
                   & Lam-Meyer               & $\mathcal{O}(T_{max} L M Q/P)$       & $\mathcal{O}(M P)$   &  this paper \\[0.5em]  \hline
Baum-Welch         & Baum-Welch              & $\mathcal{O}(T_{max} L M Q/P)$       & $\mathcal{O}(M L + P)$        & \cite{Durbin1998} \\
                   & checkpointing           & $\mathcal{O}(T_{max} L M Q \log(L/P))$ & $\mathcal{O}(M \log(L))$ &\cite{Grice1997} \\
                   & linear-memory           & $\mathcal{O}(T_{max} L M Q/P)$       & $\mathcal{O}(M)$       & \cite{Miklos2005} \\ \hline

stochastic~EM      & forward \& back-tracing  & $\mathcal{O}(T_{max} L (M + K) Q/P)$ & $\mathcal{O}(M L)$ & \cite{Bishop2006} \\
                   & Lam-Meyer               & $\mathcal{O}(T_{max} L M K Q/P)$     & $\mathcal{O}(M K P + T_{max})$ & this paper \\ \hline
                 \end{tabular}
               \end{center}
               \caption{\label{tab:overview} {\bf Theoretical computational requirements.}
                 Overview of the theoretical time and memory requirements for Viterbi
                 training, Baum-Welch training and stochastic EM~training for an HMM with $M$
                 states, a connectivity of $T_{max}$ and $Q$ free parameters. $K$ denotes the
                 number of state paths sampled in each iteration for every training sequence
                 for stochastic EM~training. The time and memory requirements below are the
                 requirements per iteration for a single training sequence of length $L$.  It
                 is up to the user to decide whether to train the $Q$ free parameters of the
                 model sequentially, i.e.\ one at a time, or in parallel in groups. The two
                 tables below cover all possibilities.
                 In the general case we are dealing with a training set $\mathcal{X} = \{X^1,
                 X^2, \ldots, X^N\}$ of $N$ sequences, where the length of training sequence
                 $X^i$ is $L^i$. If training involves the entire training set, i.e.\ all
                 training sequences simultaneously, $L$ in the formulae below needs to be
                 replaced by $\sum_{i=1}^N L_i$ for the memory requirements and by
                 $\max_i\{L_i\}$ for the time requirements. If, on the other hand, training
                 is done by considering by one training sequence at a time, $L$ in the
                 formulae below needs to be replaced by $\sum_{i=1}^N L_i$ for the time
                 requirements and by $\max_i\{L_i\}$ for the memory requirements.
               }
             \end{table}

\bigskip

   \begin{table}
     \begin{center}
     \begin{tabular}{l||r|r|r}
       CPU time (sec) per iteration      & dishonest & extended dishonest & CpG~island  \\
       & casino    & casino             & model       \\ \hline
       Baum-Welch training               & 8.85      & 5.94               & 22.22       \\ 
       stochastic EM~training $K=1$      & 5.12      & 3.42               &  5.42       \\ 
       stochastic EM~training $K=3$      & 6.02      & 4.42               & 10.30       \\ 
       stochastic EM~training $K=5$      & 7.06      & 5.38               & 14.84       \\
       Viterbi training                  & 4.42      & 2.84               &  5.00       \\ 
     \end{tabular}
     \caption{{\bf CPU time use for different models.}
       Overview of the CPU time usage in seconds per iteration for Viterbi
       training, Baum-Welch training and stochastic EM~training for the three
       different models.  For each model, we implemented each of the three training
       methods using the linear-memory algorithms for Baum-Welch training, Viterbi
       training and stochastic EM~training. The number of state paths that are
       sampled for each iteration and each training sequence in stochastic EM~training
       is denoted $K$.}
   \end{center}
   \end{table}

\end{bmcformat}
\end{document}